\documentclass[showpacs,amsmath,amssymb,reprint,prab,noeprint]{revtex4-2}

\usepackage{graphicx}
\usepackage[table]{xcolor}
\usepackage{bm}
\definecolor{prabblue}{rgb}{0.0, 0.0, 0.6}

\usepackage[
  colorlinks=true,
  linkcolor=prabblue,
  citecolor=prabblue,
  urlcolor=prabblue
]{hyperref}

\begin{document}

\title{Experimental verification of space charge saturation scaling laws \\ in high gradient photocathode RF guns}
\author{P. Denham}
\author{D. A. Garcia}
\author{A. Kulkarni}
\author{B. Schaap}
\author{Z. Liu}
\author{P. Musumeci}

\affiliation{Department of Physics and Astronomy, University of California at Los Angeles, Los Angeles, California 90095, USA}


\author{D. Filippetto}
\affiliation{Lawrence Berkeley National Laboratory, Berkeley, California 94720, USA}

\date{\today}


\begin{abstract}
We investigate the limits of photoemission yield in a high-gradient S-band radiofrequency photoinjector in the space-charge–dominated regime. Using an RF phase-scan technique, where the emitted charge is measured as a function of the phase of the RF field in the gun, we directly monitor photoemission over a range of launch fields and for different laser parameters, enabling quantitative characterization of space-charge saturation. Measurements, supported by simulations and analytic modeling, confirm the characteristic charge–field scaling laws for pancake-beams and provide the first experimental verification of cigar-regime scaling in an RF photogun. These results establish a predictive framework for identifying the onset of space-charge saturation and guide the optimization of photoinjectors for ultrafast electron diffraction, microscopy, and high-brightness light sources operating at ultra-high gradients.
\end{abstract}




\maketitle 


\section{Introduction}

Radiofrequency (RF) photoguns have become the workhorse for generating high-brightness electron beams. These beams are essential for a broad class of experiments, from radiation generation applications including short wavelength XFEL, inverse Compton scattering \cite{Bahrdt2013FirstRadiation,Jentschura2011GenerationBackscattering} and coherent THz sources ~\cite{PhysRevLett.107.204801}, to directly probing ultrafast structural dynamics in time-resolved electron microscopy and diffraction~\cite{RevModPhys.94.045004}, to injection into high-frequency advanced accelerators \cite{wu2021high,PhysRevAccelBeams.21.073401}. All of these applications impose strict requirements on the electron current density and overall bunch distribution ~\cite{RevModPhys.94.025006,RevModPhys.86.897, musumeci2018advances}. A useful metric to capture the beam quality is the peak phase space density, or beam brightness. This quantity is strongly influenced by the photoemission process and by the near-cathode dynamics, where the combination of low beam energy and proximity to the boundary makes space-charge forces particularly significant, with the potential to degrade beam quality. Naturally, one of the major pushes in electron gun technology is to maximize the initial accelerating field experienced by the particles, precisely in order to minimize the time spent by the particles in this critical region and accelerate them as quickly as possible away from the cathode to relativistic energies. 

As we increase the amount of charge in the bunch, an important limit is reached when the beam self-fields can become so large that they can fully shield the applied gun field, so that additional electrons photoemitted in these conditions are effectively turned back into the cathode, thus "saturating" the photoemission yield. Accurately predicting how much charge can be loaded into a single pulse before space charge effects saturate the extraction field is then one of the key elements (together with the initial momentum distribution) in determining the maximum brightness achievable from a photoinjector. Understanding the scaling of the saturation charge with electric field gradient is complicated by the intrinsic dependence on the laser illumination geometry and for RF photoinjectors by the time-varying nature of the RF fields and remains a subject of active research \cite{PhysRevLett.102.104801,PhysRevSTAB.17.024201}. 

For a DC current, this limit can be analytically estimated using variations of the Child--Langmuir (CL) law~\cite{PhysRevSeriesI.32.492,10.1063/5.0042355,PhysRev.21.419}. Originally derived over a century ago, CL theory finds a limit in beam current based on a one-dimensional, steady-state solution to Poisson’s equation. However, when the electrons are emitted by a pulsed photocathode laser of finite transverse dimension, the process is inherently time-dependent and 3D and the CL-predicted limit for the current density needs to be refined ~\cite{10.1063/1.4939607,PhysRevE.85.056408}. 

Scaling laws for space charge saturation in fact, depend strongly on the beam geometry and aspect ratio \cite{10.1063/1.1463065,Gunnarsson:2020bmf,PhysRevAccelBeams.24.123401}. In the so-called cigar regime where the pulse length is much longer than its transverse size, the saturation current is found to scale as $(E_0 R)^{3/2}$, where $R$ is the radial extent of the photoemitting area, echoing the form of the classic CL law \cite{PhysRevSTAB.17.024201}. In contrast, in the pancake regime, where the emission is transversely wide and temporally short, the saturation charge varies linearly with the applied field \cite{PhysRevLett.102.104801}. 
In the mean field approximation, where we can neglect the binary interactions between the particles, particle tracking simulations with self-consistent space charge calculations, such as those provided by modern, computationally optimized, PIC-based tools can be used to validate these scalings, which by now are well accepted in the community. Notably, however, there have been no direct experimental studies designed to isolate the differences between the various regimes under otherwise identical conditions (same electric field, laser, transverse profile, etc.).


In this paper, we present a detailed experimental comparison of space-charge saturation in the pancake and cigar photoemission regimes as a function of the initial accelerating field. The measurements are performed at the UCLA Pegasus S-band RF photoinjector and rely on the phase scan technique, i.e. measuring the photoemission yield as a function of the launch phase (and hence of the launch field) for various laser geometries. Test of different illumination conditions is enabled by UV photoemission from a cesium telluride (Cs\textsubscript{2}Te) photocathode with a quantum efficiency of approximately 3\%, illuminated by a 260~nm laser. This is important as the shaping methods utilized to tailor the laser pulse properties are inherently lossy, but still leave sufficient overhead in the laser energy to be able to reach the saturation regime in all cases. Excellent agreement between the experimental measurements, the theoretical predictions and the particle tracking simulations in the space-charge-dominated regime showcase the transition from the QE-dominated emission to space-charge-dominated emission for different laser pulse lengths. For the same illuminated area, it is found that significantly more charge can be extracted in the long laser pulse regime compared to the pancake case, offering practical guidance for identifying the space-charge saturation limit and optimizing beam brightness in a particular photoinjector setup. 

This paper is organized as follows. We first revisit the theoretical limits of space-charge saturation in photoemission, presenting a compact analytic derivation of the scaling laws originally obtained in Ref.~\cite{PhysRevSTAB.17.024201} and extending them to non-uniform transverse emission profiles. These results are validated in simulation with \textsc{General Particle Tracer} (GPT) using its \texttt{spacecharge3dmesh} PIC module~\cite{gpt} with the cathode boundary condition. We then focus on the main point of the work which is the experimental investigation of the space charge saturation regime: we describe the photoinjector and diagnostics setup and illustrate the phase-scan method used to quantitatively locate the onset of saturation as a function of the launch-field. We then present systematic launch-field and laser-energy scans for both pancake and cigar pulses at different spot sizes; the observed trends (e.g., $Q_{\rm sat}\!\propto\!E^{3/2}$ and $Q_{\rm sat}\!\propto\!R^{3/2}$) are found to be consistent with both the analytic scaling laws and GPT predictions. These results provide a practical procedure for selecting operating setpoints just below the onset of space-charge saturation, beyond which the emission will be affected by
virtual-cathode instabilities \cite{PhysRevA.27.1535}, characterized by photocurrent transient oscillations and a degradation of beam brightness\cite{10.1063/1.1728361}.

\section{Space charge saturation limits}

\begin{figure}
    \centering
    \includegraphics[width=0.99\linewidth]{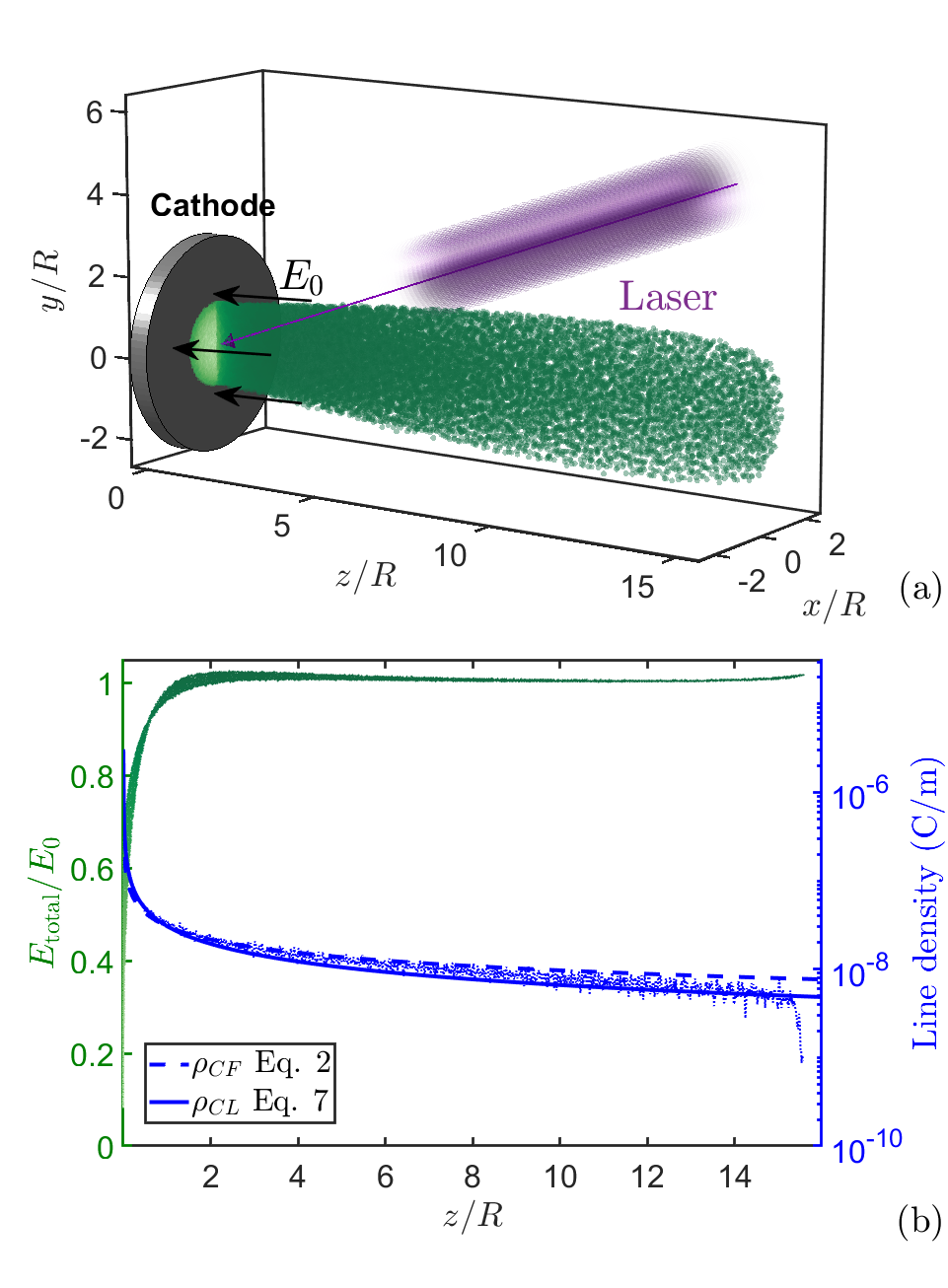}
    \caption{ (a) Diagram based on GPT spacecharge3dmesh PIC simulation of space charge limited photoemission model and relevant parameters. (b) The axial dependence of the electric field within the beam, as well as the charge density that produces it. Analytical expressions from Eq. \ref{eq:rhocf} and Eq. \ref{eq:rhocl} are also plotted for the density.}
    \label{fig:PIC_sim} 
\end{figure}

A schematic visualization of the photoemission process is shown in the top panel of Fig.~\ref{fig:PIC_sim}. The purple beam represents the photocathode drive laser, and a longitudinal bisection of the electron bunch in the $xz$-plane is shown at the end of the laser pulse. To be consistent with the experimental parameters used later, this example is generated with GPT using a static external field $E_0=50$~MV/m, a disk emitter of radius $R=50~\mu$m, a uniform temporal profile of duration $T=13.2$~ps, and total charge $Q=12$~pC. The electrons, accelerated primarily along $z$, expand into a nearly cylindrical distribution of radius $R$. The GPT Poisson solver evaluates the total field as the sum of the applied gun field, image charges, and the bunch’s own space charge. The bottom panel of Fig.~\ref{fig:PIC_sim} shows the corresponding on-axis field distribution together with the axial charge density.

Although the simulation parameters were chosen to match those used in the experiment, a few general features apply to all pulsed photoemission systems. There is a region close to the cathode of axial extent approximately equal to the beam transverse size $R$ where the space charge field is large enough to suppress the extraction field. If the beam axial extent at the end of the laser pulse $\tfrac{1}{2}(eE_0/m)T^2$ is much greater than $R$ (like in our simulation case), the emission is in the cigar regime. In contrast, when the laser pulse is short enough that the bunch length at the end of the emission is comparable to or smaller than the beam radius, the electrons form a wide, thin sheet characteristic of the pancake regime. The degree of field suppression depends on the extracted charge. When the self-field locally cancels the applied field at the cathode, emission saturates and additional photoelectrons generated in that region are driven back into the surface, even if other parts of the beam spot may continue to emit.

In this section we review the conditions under which this threshold behavior occurs and revisit the analytical scaling laws that quantify the saturation charge $Q_{\mathrm{sat}}$, defined as the minimum charge required to cancel the applied field at the cathode center.

A rigorous treatment would require solving Maxwell’s equations (e.g., Jefimenko’s generalizations of the Coulomb law \cite{jackson3}) to capture the time-dependent, self-consistent charge and field evolution. Additional complications arise from thermal velocity spread and relativistic dynamics, which demand a full kinetic description. Such an approach is beyond the scope of this work. Instead, we adopt heuristic estimates guided by PIC simulations in the classical regime, valid when electron initial velocities can be neglected and the laser pulse duration satisfies
$\frac{R}{c} \ll T \ll \frac{mc}{eE_0}$.

Within these approximations, the space charge field at the cathode plane can be written in the Coulomb electrostatic limit as
\begin{equation}
    \mathbf{E}_{sc}(\mathbf{r}_\perp, z=0) =
    \frac{1}{2\pi\epsilon_0}
    \int \rho(\mathbf{r}',z') \,
    \frac{\mathbf{r}_\perp - \mathbf{r}'}{|\mathbf{r}_\perp - \mathbf{r}'|^3} \, d^3r',
    \label{Eq:scfield_cathode}
\end{equation}
where $\mathbf{r}_\perp$ is the transverse offset from the cathode center, $\rho(\mathbf{r}',z')$ is the charge density, $\epsilon_0$ is the vacuum permittivity, and the integral extends over the beam volume $V_b$. This form is valid when retardation can be neglected, i.e. $|\mathbf{r}_\perp| \ll cT$. For longer pulses or larger transverse offsets, retarded field effects would have to be included. The prefactor of 2 accounts for the conducting cathode boundary via the method of images.

Following the discussion in ~\cite{10.1063/1.5063888,PhysRevLett.87.278301}, we neglect transverse dynamics near the cathode and the emitted beam is taken to have the same spatial distribution as the laser spot. Temporally, we assume a uniform laser profile of duration $T$, so the total charge is $Q=I_0T$ with peak current $I_0$. For a cylindrically symmetric profile, the charge density can be written as $\rho(r,z)= J(r)/v(z)$, where $v(z)$ is the longitudinal velocity and $J(r)$ is the current density. The latter is written as $J(r)=J_0 f(r)$ in terms of a normalized transverse distribution function $f(r)$ (i.e. $\max f=1$) and the peak current density $J_0=I_0/A_{\mathrm{eff}}$ which depends on the peak current and the effective emission area $A_{\mathrm{eff}}=\int_0^\infty 2\pi r f(r)\,dr$.

To evaluate field integrals, we introduce a characteristic transverse size $R=\sqrt{A_{\mathrm{eff}}/\pi}$ (equal to the beam radius for a uniform profile) to define normalized longitudinal and radial coordinates $\xi = z/R$ and $u=r/R$.


In the absence of space charge, the velocity field is found by solving the equation of motion for an electron in a constant field, $v(z)=\sqrt{2 e E_0 z / m}$, resulting in a density proportional to $z^{-1/2}$, which can be expressed as:
\begin{equation}
\rho_{CF}= 
\sqrt{\frac{m}{2eE_0z}}
     J_0 f(r),
     \label{eq:rhocf}
\end{equation}
where the $CF$ subscript is used to denote this constant field approximation. For longer pulses with higher beam charge, this ballistic approximation is too crude, and as electrons spill into the vacuum, the velocity field gets modified towards a configuration resembling the Child-Langmuir classical profile that is $v(z)\propto z^{2/3}$ (so that $\rho_{CL}\propto z^{-2/3}$).

\begin{figure*}
    \centering
    \includegraphics[width=0.375\linewidth]{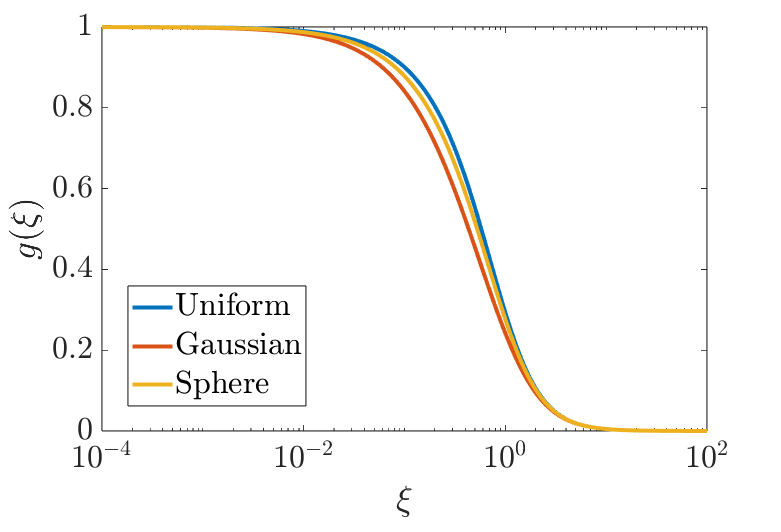}(a)
    \includegraphics[width=0.4\linewidth]{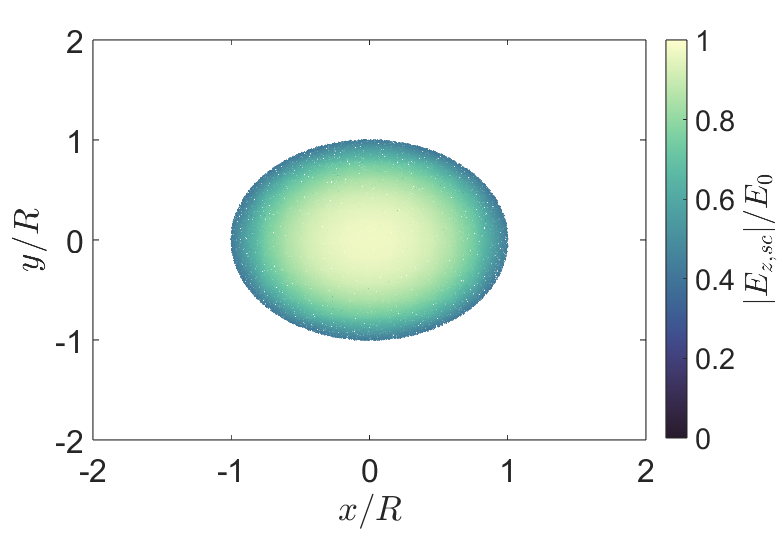}(b)
    \includegraphics[width=0.4\linewidth]{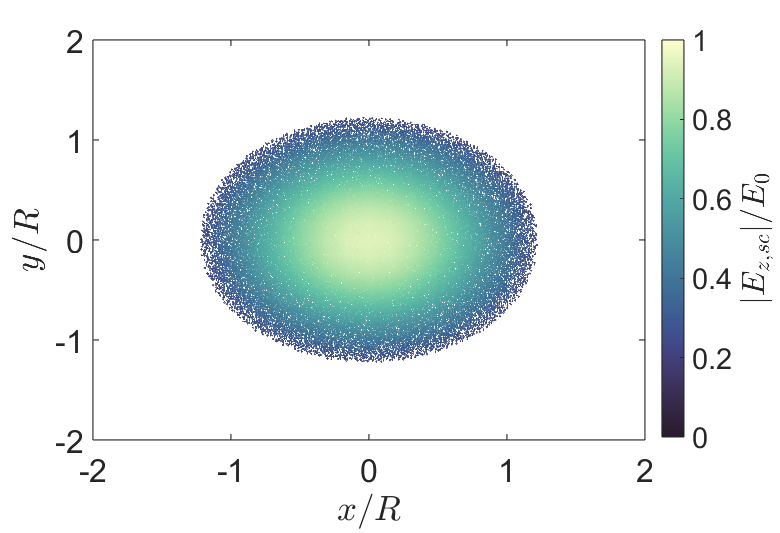}(c)
    \includegraphics[width=0.4\linewidth]{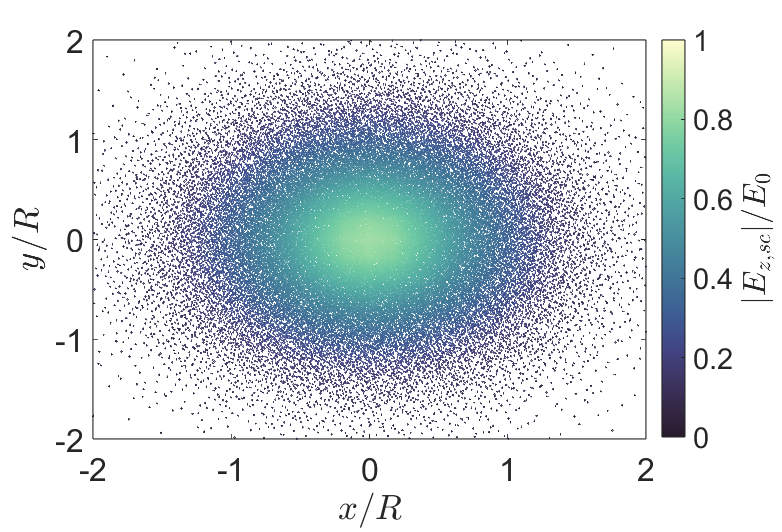}(d)
    \caption{(a) Plot of $g(\xi)$ for the uniform, Gaussian, and sphere distributions with the $\xi$-axis in log scale. Example cigar regime cathode space charge field distribution in the x-y plane for (b) a Uniform emission profile, (c) a spherical emission profile, and (d) a Gaussian emission profile; with all having the same effective area.}
    \label{fig:cathodefield}
\end{figure*}

After substituting the density $\rho_{CF}$ into Eq. \ref{Eq:scfield_cathode} and carrying out the integral over the beam area, the space charge field at the cathode center can be expressed as a single integral in the variable $\xi$, where the final limit of integration is the final aspect ratio $\xi_f = \frac{e E_0 T^2}{2mR}$
\begin{equation}
E_{sc,CF} = \frac{Q}{\epsilon_0 A_\text{eff}} \sqrt{\frac{mR}{2eE_0T^2}} \int_0^{\xi_f} \frac{1}{\sqrt{\xi}} g(\xi) \, d\xi.
\label{eq:ballisticint}
\end{equation}
and the function $g(\xi)$ depends on the initial transverse emission profile and is found integrating over the beam cross section $g(\xi) = \int \frac{f(u)\xi}{(u^2+\xi^2)^{3/2}} u du$. The resulting functional form of $g(\xi)$ for the uniform, Gaussian and spherical profiles are listed in Table \ref{tab:distributions}, and plotted in Fig.~\ref{fig:cathodefield}(a), where it can be seen they are very close in shape.
 
Physically $g(\xi)$ can be interpreted as the normalized on axis electric field of a disk with a surface charge density distribution proportional to $f(r)$. For $\xi$ approaching 0 (i.e. at distances from the disk much smaller than the disk extent), we do expect to recover the field of an infinitely wide plane and so $g(\xi)\rightarrow 1$.  

\begin{table}[t]
\centering
\small
\renewcommand{\arraystretch}{1.0}
\setlength{\tabcolsep}{5.3pt} 
\begin{tabular}{lccc}
\hline\hline
\textbf{Dist.} & $f(r)$ & $A_{\mathrm{eff}}$ & $g(\xi)$ \\
\hline
\textbf{Uniform} &
$\begin{cases}
1, & r < R \\
0, & r \ge R
\end{cases}$ &
$\pi R^2$ &
$1 - \dfrac{\xi}{\sqrt{\xi^2 + 1}}$ \\[0.6em]
\textbf{Gaussian} &
$\exp\!\left(-\dfrac{r^2}{2\sigma^2}\right)$ &
$2\pi\sigma^2$ &
$1 - \sqrt{\pi}\,\xi\, e^{\xi^2}\operatorname{erfc}(\xi)$ \\[0.6em]
\textbf{Sphere} &
$\sqrt{1 - r^2/a^2}$ &
$\dfrac{2}{3}\pi a^2$ &
$1 - \sqrt{\dfrac{2}{3}}\,\xi\, \tan^{\!-1}\!\!\left(\!\frac{\sqrt{\tfrac{3}{2}}}{\xi}\right)$ \\
\hline\hline
\end{tabular}
\caption{Relevant distributions with associated effective areas and corresponding $g(\xi)$.}
\label{tab:distributions}
\end{table}

We can then use this general expression and set the space-charge field at the cathode center equal in magnitude to the applied field to solve for the saturation charge. For example, in the pancake limit; when \( T \to 0 \) we have
\begin{equation}
\begin{aligned}
E_0 &=  \frac{Q_\text{sat}(T \rightarrow 0)}{2\epsilon_0 A_\text{eff}} 
       \lim_{\xi_f \to 0} \frac{1}{\sqrt{\xi_f}} \int_0^{\xi_f}  \frac{1}{\sqrt{\xi}} g(\xi)\, d\xi \\
    &= \frac{Q_\text{sat}(T \rightarrow 0)}{\epsilon_0 A_\text{eff}}.
\end{aligned}
\end{equation}
after applying L'H\^opital's rule to calculate the limit, thus recovering the pancake regime result \cite{PhysRevLett.102.104801},
\begin{equation}
Q_\text{sat}(T \rightarrow 0) = \epsilon_0 A_\text{eff} E_0,
\label{eq:pancake}
\end{equation}
where the saturation charge or maximum charge extractable from a photoinjector is directly proportional to the field and effective emitter area.

If we now consider long laser pulses where the duration $T\gg\sqrt{2mR/eE_0}$ or equivalently $\xi_f \gg 1$, it is convenient to recast Eq. \ref{eq:ballisticint} in terms of the saturation peak current $I_0$ for which the space charge field fully cancels the cathode extraction field. We perform the integral for all cylindrically symmetric transverse profiles, yielding
\begin{equation}
\begin{aligned}
    E_{sc,CF}= 2\sqrt{2}\frac{I_0}{I_a} \left(\frac{mc^2}{e R}\right)^{\frac{3}{2}}\frac{1}{\sqrt{E_0}} \int_0^{\infty} \frac{1}{\sqrt{\xi}} g(\xi) \, d\xi
\end{aligned}
\label{eq:z12}
\end{equation}
where $I_a=4\pi\epsilon_0mc^3/e=17kA$ is the Alfven current. The upper limit of integration can be extended to infinity because the main contributions to the field come from charges within a few units from the cathode, hence adding in the sources at large $\xi$ has little impact on the measure, as seen explicitly in Fig. \ref{fig:cathodefield}(a). This is also supported by looking at the density profile from the PIC simulation in Fig. \ref{fig:PIC_sim}(b). For example, in the uniform illumination case, the value of the dimensionless integral
\(\int_0^{\infty} \xi^{-1/2}\!\left(1 - \xi/\sqrt{\xi^2+1}\right)\, d\xi=2 \Gamma(3/4)^2/\sqrt{\pi }\approx1.694\). Hence, an upper bound on the limiting current can be expressed as $I_{CF}=0.208I_a\left(\frac{e E_0 R}{m c^2}\right)^{3/2}$.


\begin{figure*}
    \centering
    \includegraphics[width=0.45\linewidth]{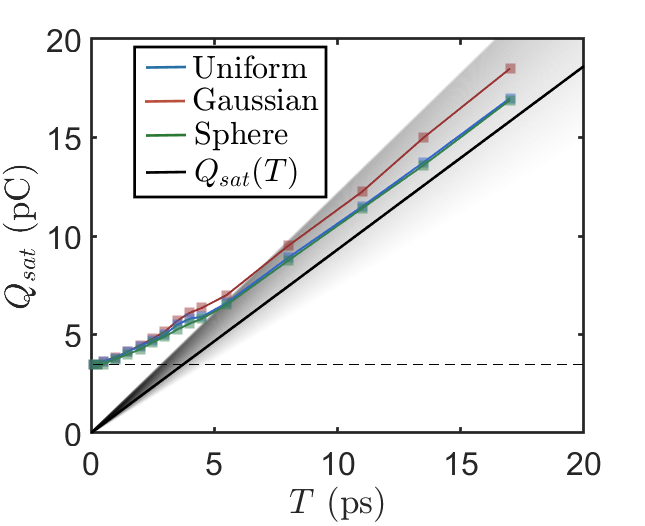}(a)
    \includegraphics[width=0.45\linewidth]{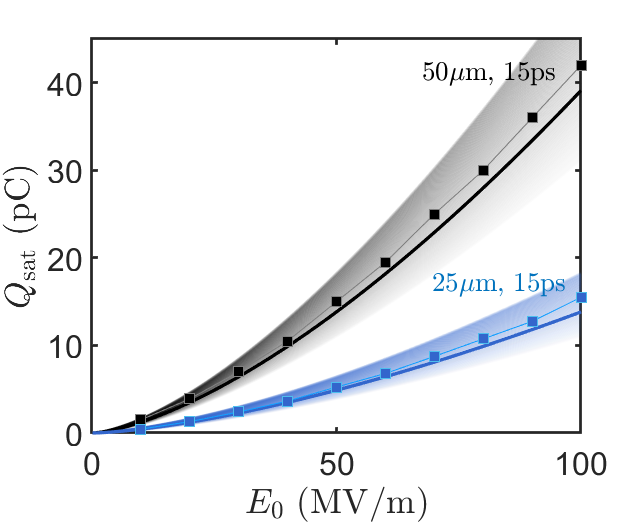}(b)
    \caption{(a) Charge saturation versus pulse duration for different transverse profiles, compared at equivalent effective area $\sqrt{2/3}\,a=\sqrt{2}\,\sigma=R=50\,\mu\mathrm{m}$ and $E_0=50$~MV/m. Data points from GPT simulations are shown for Gaussian (red), uniform (blue), and spherical (green) profiles, marking the input charge at which the on-axis cathode field vanishes. The horizontal dashed line indicates the pancake saturation limit; the solid black line shows the uniform-profile prediction; and the shaded band denotes the critical saturation interval. (b) Saturation charge versus peak field $E_0$ for $T=15$~ps. Theory curves from Eq.~\ref{eq:cigar} are shown as solid lines for $R=50\,\mu\mathrm{m}$ (black) and $R=25\,\mu\mathrm{m}$ (blue) assuming uniform transvere profiles (i.e. $C_{dist} = 1)$. The shaded bands again denote the saturation interval. GPT simulation points for the uniform case are overlaid.}

    \label{fig:pulse_length_dependence}
\end{figure*}
At this point, it is also worth noting that, as soon as the photocurrent is driven into the vacuum, it will begin screening the external field, thus modifying the electron velocity distribution and undermining the 'constant field' density approximation of Eq. \ref{eq:ballisticint}. At the limit, we can expect near steady state conditions where the space-charge field reconfigures the axial density into a Child-Langmuir-like profile \cite{10.1063/1.5063888}. Accordingly, we can adapt the formalism by incorporating a 1D Child-Langmuir density with a suitably scaled emission profile, from which we obtain:
\begin{equation}
    \rho_{\rm CL}
  = 
  \left(\frac{2}{9}\frac{\epsilon_0 m}{e}\right)^{1/3}
     \left(\frac{J_0}{z}\right)^{2/3}
     f(r)
     \label{eq:rhocl}
\end{equation}


Replacing the constant-field density by the CL profile $\rho_{\mathrm{CL}}$ yields the more realistic limiting form:
\begin{equation}
E_{sc,CL}=\sqrt[3]{\tfrac{32}{9}}\left(\frac{I_0}{I_a}\right)^{2/3}\frac{mc^2}{eR}\int_0^{\infty} \frac{1}{\xi^{2/3}}g(\xi)\, d\xi
\end{equation}
In the case of uniform illumination, the integral evaluates as \(\int_0^{\infty} \xi^{-\tfrac{2}{3}} \left(1 - \xi/\sqrt{\xi^2+1}\right) \, d\xi=3 \Gamma(2/3) \Gamma(5/6)/\sqrt{\pi }\approx2.59\).
Just as before, we equate to the applied field strength and find the space charge limiting current for the CL density is given by $I_{CL}=0.127 I_a\left(\frac{e E_0 R}{m c^2}\right)^{3/2}$.

Notably in either the CF or CL case, the two density profiles lead to the same $(E_0R)^{3/2}$ scaling in the cigar regime. Note that in Fig. \ref{fig:PIC_sim}(b) neither of the expressions capture the full behavior of the charge density over the entire beam extent. The actual saturation current is expected to lie in an interval between these limiting bounds (i.e. $I_{CL}\lesssim I_{sat}\lesssim I_{CF}$), and thus share the same scaling. The saturation charge in this regime is very closely approximated by the formula originally given in \cite{PhysRevSTAB.17.024201}
\begin{equation}
    Q_{\mathrm{sat}}(T)\simeq C_{\mathrm{dist}}\,I_a\,T\,\frac{\sqrt{2}}{9}
    \left(\frac{e E_0 R}{m c^2}\right)^{3/2},
    \label{eq:cigar}
\end{equation}
where $C_{\mathrm{dist}}$ is a order of unit prefactor which accounts for the small dependence on the transverse emission profile. Specifically, $C_{\mathrm{dist}}$ is normalized to unity for the uniform case; and by performing the integrals over $g(\xi)$ we numerically calculate $C_{\mathrm{dist}}\!\approx\!1.12$ for the Gaussian and $C_{\mathrm{dist}}\!\approx\!1.04$ for the spherical distribution.

Varying cathode transverse illumination profiles result in different electric field distributions on the cathode. However, for distributions with equivalent effective areas (shown in Fig.~\ref{fig:cathodefield}b–d), the onset of an extinguished extraction field at the origin should occur for nearly the same input charge based on the analytical results, and it is found that for equal effective area, the discrepancy between the onset charge saturation is only a few percent. This trend continues into modest pulse lengths, as shown in Fig.~\ref{fig:pulse_length_dependence}(a), where the saturation onset is plotted against the incident laser pulse length. The data points are GPT simulations for uniform (blue), Gaussian (red), and spherical (green) profiles, representing the lowest simulation input charge for which the field at the cathode center is fully canceled. The solid black line is Eq.~(\ref{eq:cigar}), while the opaque band represents the interval where, depending on the actual velocity field profile, saturation is expected to occur. For very short bunch lengths the saturation onset occurs at the pancake aspect ratio limit given by Eq. \ref{eq:pancake} represented by the horizontal dashed line. For intermediate length laser pulses, the simulated results connect smoothly the two asymptotic limits discussed earlier, reflecting a bunch profile that transitions between the cigar and pancake regimes. In this case, the saturation charge could still be described by Eq.~(\ref{eq:z12}) without taking the limit in the integral.

The scaling of saturation with emission radius implies that doubling the radius increases the saturation charge by a factor of $2^{3/2}$. This characteristic is numerically validated in Fig.~\ref{fig:pulse_length_dependence}(b), where charge saturation is plotted as a function of peak field $E_0$ for a fixed pulse duration of $T = 15~\mathrm{ps}$. Theoretical predictions are shown for two emission spot sizes, $R = 50~\mu\mathrm{m}$ (black) and $R = 25~\mu\mathrm{m}$ (blue), assuming a uniform transverse distribution. The solid lines correspond to Eq.~\ref{eq:cigar}, while the shaded bands indicate the critical saturation interval defined by the same range of multiplicative prefactors used in Fig.~\ref{fig:pulse_length_dependence}(a). Data points from GPT simulations are overlaid, confirming that the saturation point shifts by exactly a factor of $2^{3/2}$ across all extraction fields when the spot radius is doubled. An experimental verification of this scaling is presented in a later section.  

\section{Experimental Setup}

\begin{figure*}
    \centering
    \includegraphics[width=0.8\linewidth]{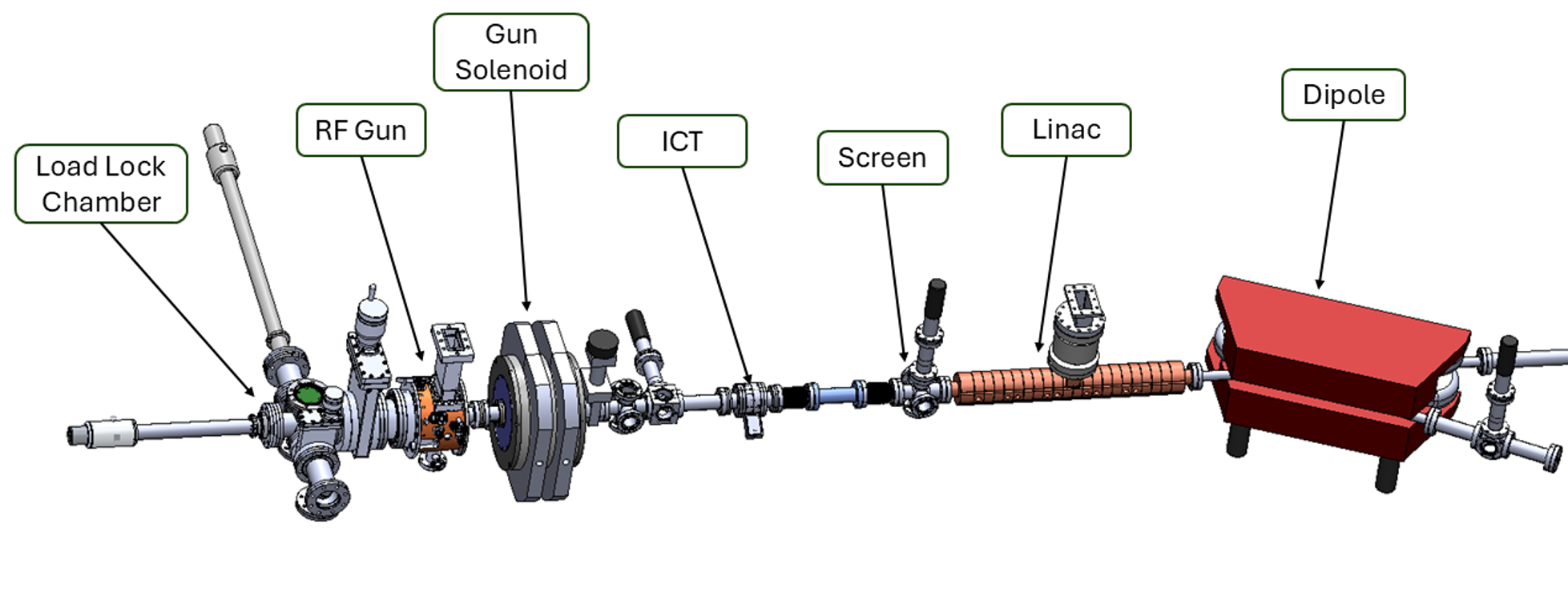}(a)
    \includegraphics[width=0.45\linewidth]{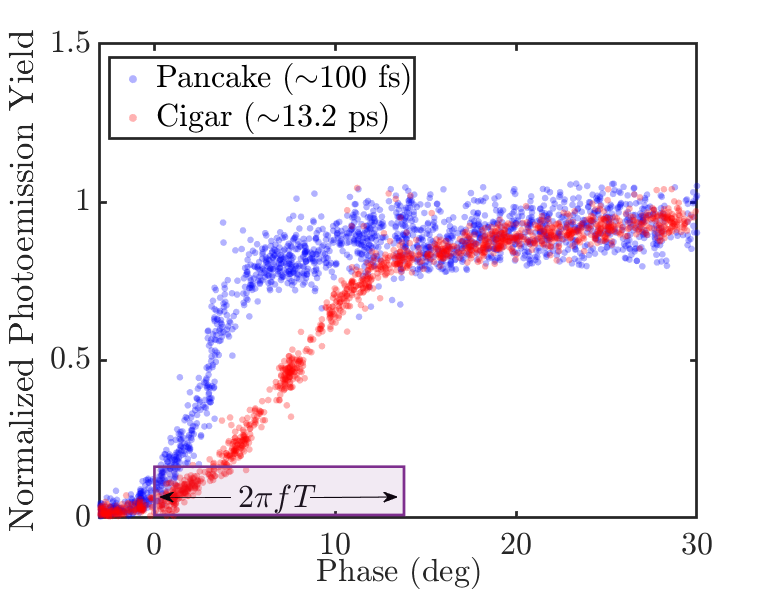}(b)
    \includegraphics[width=0.45\linewidth]{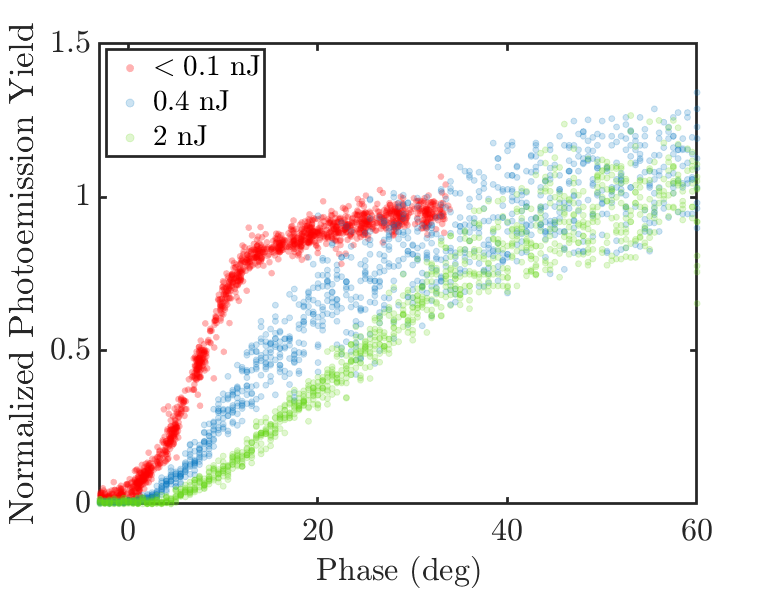}(c)
        \caption{(a) Relevant section of the Pegasus beamline used for the measurement. (b) QE-limited phase scans for short-pulse (pancake) and long-pulse (cigar) illumination. The pancake rise time is broadened beyond the expected 0.1 deg by the intrinsic longitudinal velocity spread and finite photocathode response, while the cigar trace spans $\sim13^{\circ}$, matching the stacked-beamlet laser pulse duration. (c) Phase scans at increasing laser energy, showing rising edges that extend beyond $13^{\circ}$ as space-charge–limited emission sets in.}

    \label{fig:experiment}
\end{figure*}

\begin{figure}
    \centering
      \includegraphics[width=0.95\linewidth]{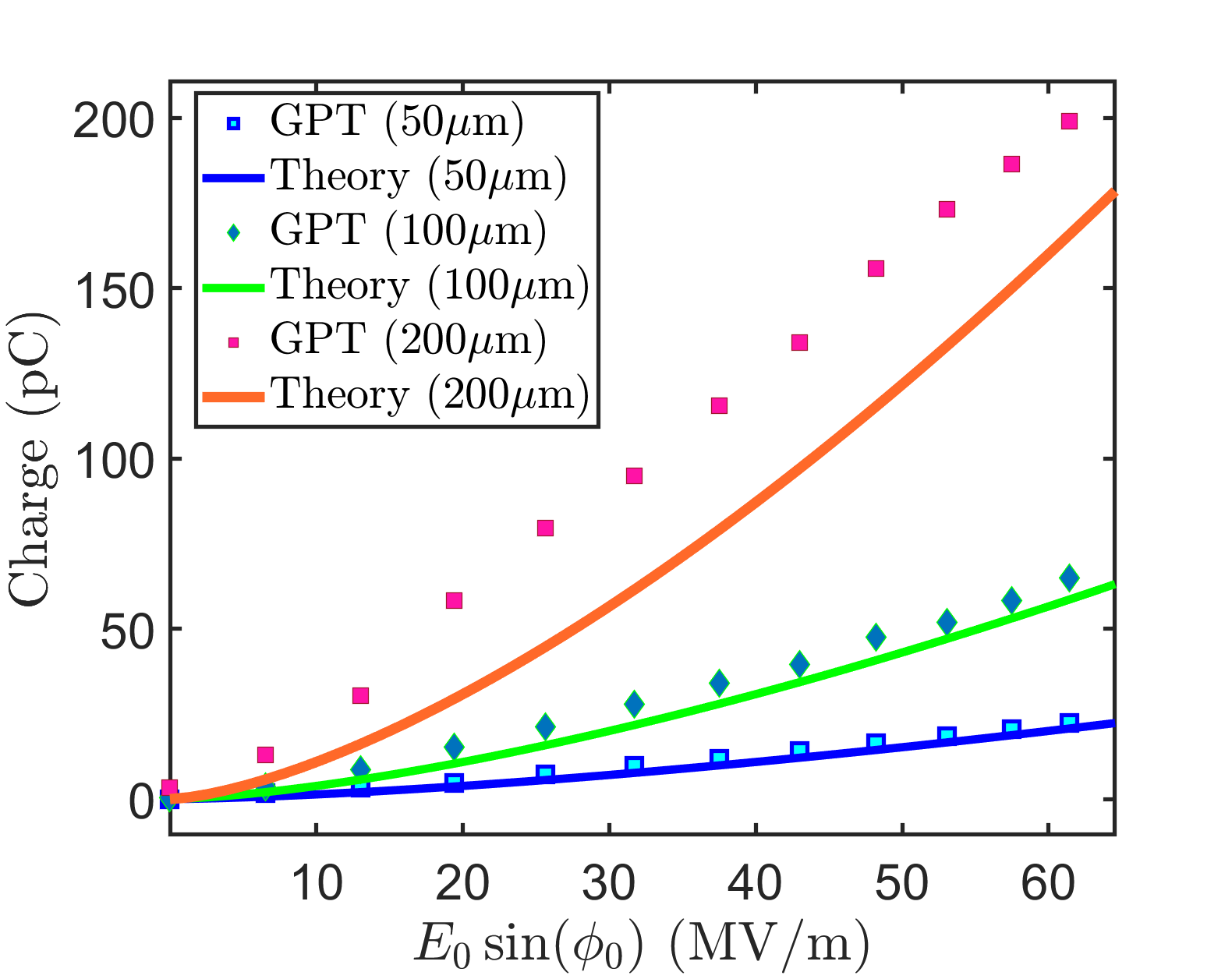}(a)     \includegraphics[width=0.9\linewidth]{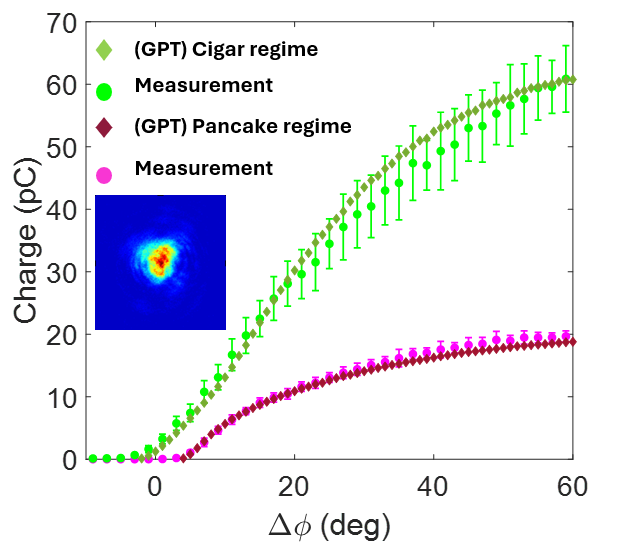}(b)
\caption{(a) Simulated phase scans (ideal uniform disks) for varying radii with corresponding \(Q_{\rm sat}\) curves parameterized by \(E_0\sin\phi_0\).
(b) Measured phase scans (pancake and cigar) with GPT overlays seeded by the measured VCC bitmap; same QE–phase calibration and return-to-cathode removal.}
    \label{fig:method_check}
\end{figure}

The experiment was conducted at the Pegasus beamline at UCLA. A schematic of the relevant section is shown in Fig.~\ref{fig:experiment}(a), which includes the load-lock system that enables operation with air-sensitive high QE Cs$_2$Te cathodes, the 1.6-cell S-band RF gun, and a focusing solenoid to provide full beam collection across the entire range of launch phases. An integrated current transformer (Turbo-ICT from Bergoz Instrumentation) is installed on the beamline for charge measurement with sub-pC resolution \cite{stulle2015turbo}. The beamline also includes an intermediate YAG screen to monitor the beam spot and check for losses and clipping, as well as an RF linac and a dipole for energy measurements that allow for accurate peak field calibration in the gun \cite{alesini:pegasus}


The laser delivery system offers significant flexibility on the parameters of the 260 nm pulse illuminating the cathode. A 75 cm focal length lens is used to image onto the cathode plane a remotely driven iris aperture, enabling spot size control. The transverse profile of the laser is recorded continuously by a camera looking at the reflection from a pick up mirror on the UV line virtual cathode camera (VCC). A short laser pulse (100 fs FWHM) was used to test the charge yield scaling in the pancake regime. For generating a cigar-shaped electron beam, instead, we employed a pulse-stacking technique using a series of birefringent $\alpha$-BBO crystals. Four crystals with descending lengths of 8~mm, 4~mm, 2~mm, and 1~mm were used to stretch, split and delay the UV laser pulse, resulting in an nearly uniform total pulse duration of approximately 13.2~ps (about 13 degrees of RF phase) \cite{musumeci2010capturing}. 


The main experimental technique employed in this study is the measurement of the beam charge as a function of the launch phase in the RF field of the gun. In practice, the scan is performed by continuously varying the RF phase while recording the charge collected downstream using the current transformer. This is a standard measurement technique in RF photoinjectors, often used to identify the working phase and verify the alignment of the laser on the cathode. Phase scans have also been used to measure the laser pulse length or diagnose the field enhancement factor (or Schottky effect) of a photocathode \cite{article_prat}. Examples of phase scans recorded at Pegasus are shown in Fig. \ref{fig:experiment}(b) and (c) respectively. In Fig.~\ref{fig:experiment}(b) we show two RF phase scans for the pancake and cigar cases. On the horizontal axis, zero phase is defined as the point where charge is first detected on the ICT, corresponding to the arrival of the laser pulse head at the cathode. The scans were performed at high excess energy and with very low laser fluence to suppress space-charge effects and are normalized to unity to facilitate their comparison. In this QE-limited regime, the rise time reflects the laser pulse duration. The knee of each curve marks the end of the laser pulse. For the short-pulse (pancake) case, the rise spans only a few degrees (less than 2 ps) as is broadened by the intrinsic spread in initial longitudinal velocities and the finite response time of the photocathode. In contrast, the long-pulse (cigar) trace extends over $\sim13^{\circ}$, in agreement with the stacked beamlet laser duration when converted to RF phase, as indicated by the shaded region.

When the laser energy is increased, the shape of the phase scan curves changes significantly. An example for the cigar-shaped illumination is shown in Fig.~\ref{fig:experiment}(c), where the scans have also been normalized to highlight the altered behavior. The rising edge of the charge–phase dependence now extends well beyond $13^{\circ}$, not because of a longer laser pulse but due to the transition from the QE-dominated regime into space-charge–limited emission.

The mechanism can be understood by considering the dependence of the launch field on the RF phase $E(\phi)=E_0\sin\phi$ where $E_0$ is the peak RF field in the gun (to be distinguished from the static DC field used in the theoretical model). At early injection phases, the effective accelerating field at the cathode is small and can be fully canceled by the space-charge field of the emitted electrons, leading to saturation of the emission. As the phase increases, the external field strengthens, allowing additional charge to be released. For the highest laser fluences, full emission is reached only when the launch phase exceeds 40 degrees; at smaller phases, even though the cathode has absorbed all the laser energy,
a large fraction of the photo-excited electrons are turned back into the cathode and cannot escape. 

This behavior provides a direct route to experimentally mapping the saturation surface 
$Q_{\rm sat}(E,R,T)$. By varying laser parameters such as spot size and pulse duration, one can adjust the beam aspect ratio while avoiding virtual-cathode formation and identify the operating point for maximum extractable charge.

In practice, two complementary data collection methods provide equivalent access to the saturation behavior. One can either vary the laser pulse energy at fixed injection phase (i.e., fixed external field) to determine $Q_{\mathrm{sat}}$, or, at constant laser energy, sweep the RF phase and remap the abscissa to the corresponding field $E = E_0 \sin \phi$. Both procedures reveal the same dependence of the emitted charge on the applied field and can therefore be used interchangeably. In the following, we adopt the phase-scan approach, which offers a more direct visualization of $Q_{\mathrm{sat}}(E)$.


\subsection*{Modeling and simulations}

Before directly comparing the analytical formulas developed in the previous section with the experimental data, it is necessary to refine the simulation framework for space-charge–limited emission in the RF photoinjector.

In the General Particle Tracer (GPT) simulations, the \texttt{spacecharge3Dmesh("Cathode")} solver was employed to enforce the conducting boundary at the cathode. Macroparticles returning to $z=0$ were explicitly removed from the simulation. The simulations include the RF field map of the gun, which has been validated against experimental measurements \cite{PhysRevLett.118.154802,schaap:2025}.


As a first step, we verified that the characteristic shapes of the RF phase scans predicted by the theory are reproduced qualitatively and quantitatively by GPT. Figure~\ref{fig:method_check}(a) compares the simulation results for ideal uniform-disk sources of varying radii with the analytical predictions of $Q_{\rm sat}\!\left(E_0\sin\phi\right)$ from Eq.~\ref{eq:cigar}. For small spot sizes the agreement is excellent.  As the spot radius increases, however, the assumption of a strongly elongated “cigar” aspect ratio breaks down, resulting, as expected, in a deviation from the asymptotic limit predicted by Eq. \ref{eq:cigar}. Importantly, this comparison demonstrates that phase scans can indeed be used to retrieve the saturation charge as a function of field, and thus can work as a direct probe of space-charge–limited emission in RF guns.


Figure~\ref{fig:method_check}(b) extends the comparison to the experimentally measured phase scans (both “pancake” and “cigar” cases) using GPT simulations seeded with the measured transverse laser distribution from the VCC. Error bars indicate the shot-to-shot standard deviation over 20 shots collected at each phase setting.

Note that even after the laser pulse has fully illuminated the cathode, the measured charge continues to increase with phase. This behavior is well known and due to the variation of the QE resulting from the lowering of the work function induced by the Schottky effect ~\cite{article_prat,schmerge2006rf}. Near the photoemission threshold, the quantum efficiency (QE) scales quadratically with the excess photon energy: 
\begin{equation}
QE(\phi) = k\!\left[h\nu-\phi_{\rm eff}(\phi)\right]^2
\label{eq:qe}
    \end{equation}
where $h\nu$ is the photon energy and the effective work function depends on the instantaneous field as
\begin{equation}
\phi_{\rm eff}(\phi) = \phi_w-\kappa\,\sqrt{E_0\sin\phi}.
\end{equation}
with $\kappa=0.037947\,\mathrm{eV}\,(\mathrm{MV/m})^{-1/2}$. 

By fitting the phase scans in the QE dominated regime one can extract both the proportionality constant and the effective work function yielding $k=0.1121~\text{eV}^{-2}$ and $\phi_w=4.447~\text{eV}$~\cite{PhysRevSTAB.12.074201}. Note that the fitted work function differs slightly from the typical literature value of 3.5 eV \cite{pierce2021beam}, likely reflecting variations in the specific preparation of our Cs–Te cathode. This Schottky-based enhancement of the QE needs to be hard-coded in the GPT particle tracking simulation in order to obtain the agreement with the data shown in Fig. \ref{fig:method_check}. 

\section{Results: characterization of space charge saturation in different regimes}
\begin{figure*}
   \includegraphics[width=0.4\linewidth]{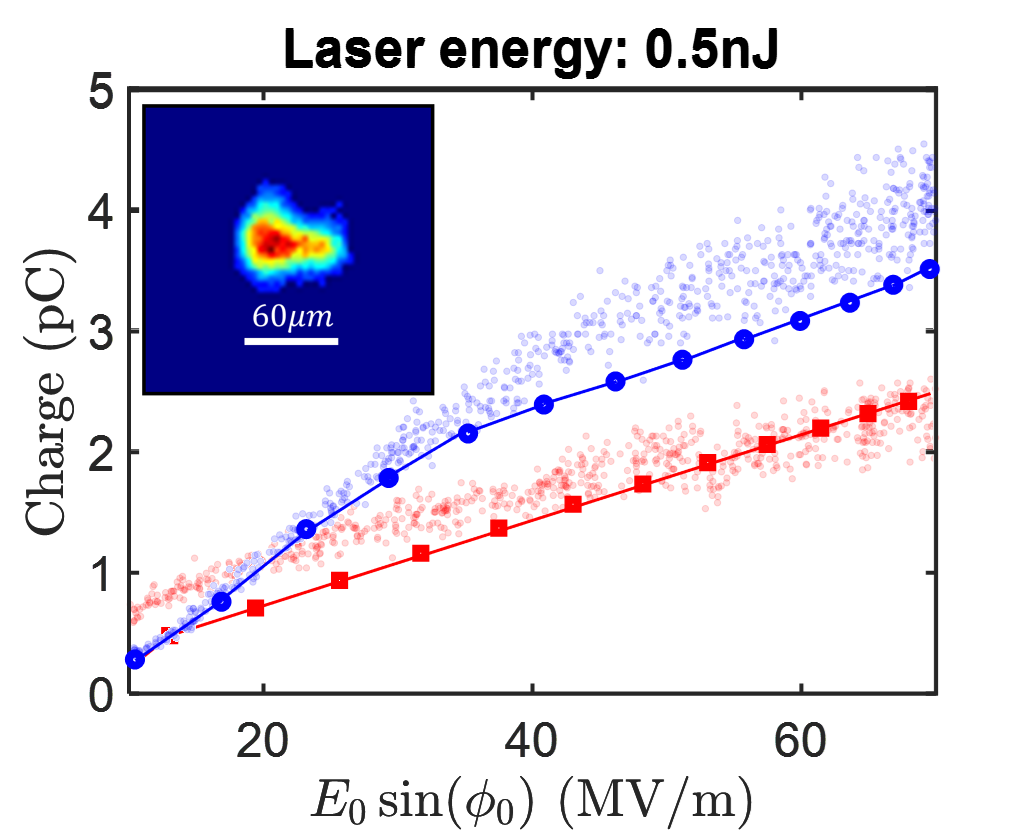}(a)
    \includegraphics[width=0.4\linewidth]{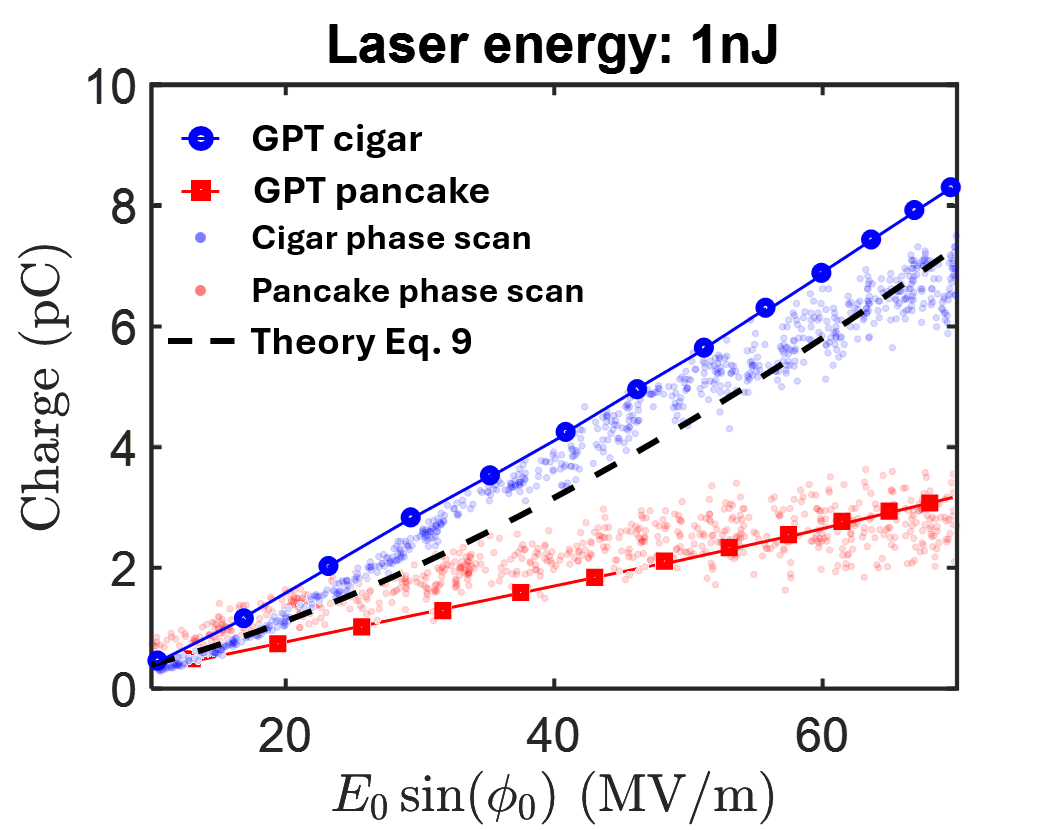}(b)
\includegraphics[width=0.4\linewidth]{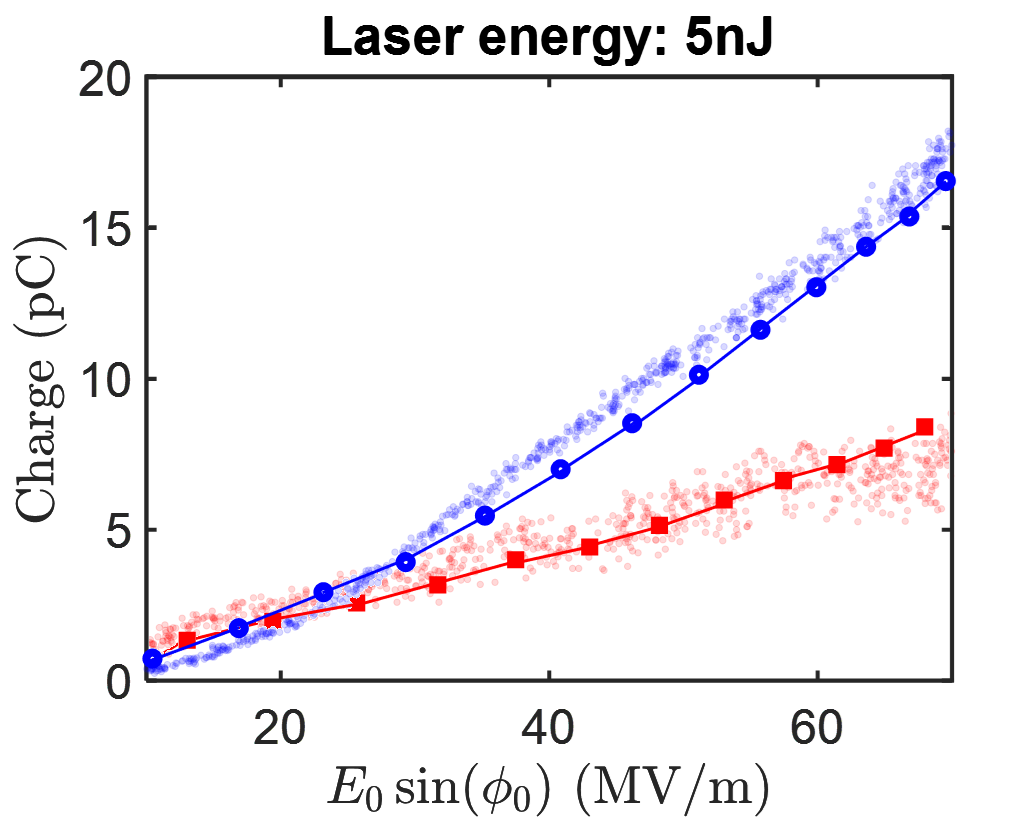}(c)
\includegraphics[width=0.425\linewidth]{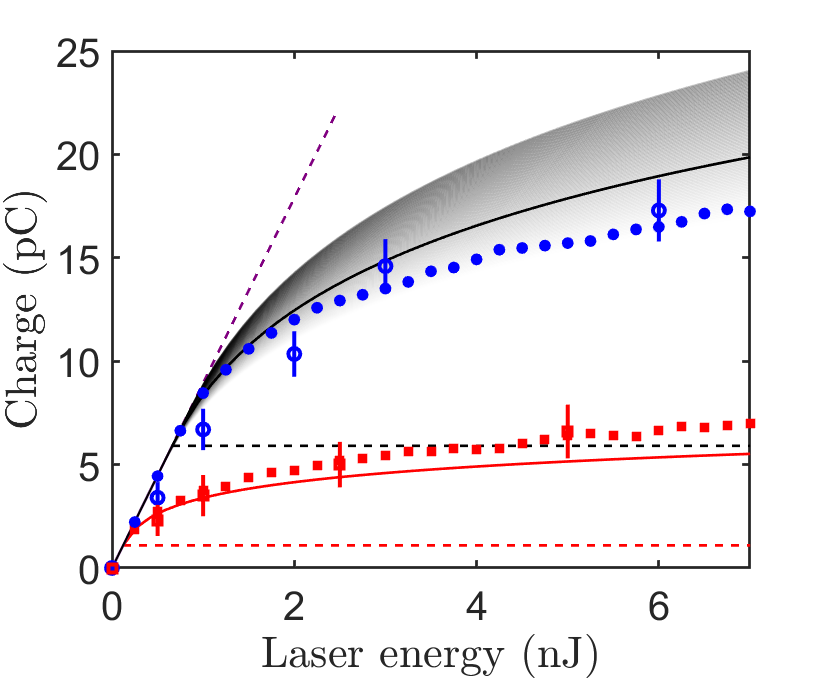}(d)

\caption{(a–c) Phase-scan measurements at three laser pulse energies comparing the pancake (red) and cigar (blue) regimes, plotted versus launch field with GPT simulations overlaid. The VCC laser profile used in the simulations is shown as an inset in~(a). (a)~At 0.5~nJ, the pancake case is fully saturated while the cigar case is only partially saturated. (b)~At 1~nJ, the pancake remains in saturation and the cigar follows the $E^{3/2}$ scaling up to $\sim$70~MV/m. (c)~At 5~nJ, both regimes are beyond saturation, with the cigar trace exceeding the theoretical curve in (b). (d)~Extracted charge versus laser energy at a fixed field of 60~MV/m. Horizontal dashed lines mark the saturation charge predicted by Eq.~(\ref{eq:pancake}) and Eq.~(\ref{eq:cigar}) for the pancake and cigar regimes, respectively. The solid red and black curves correspond to Eq.~(\ref{eq:tails}) in the pancake and cigar regimes, respectively. Squares and circles with error bars are experimental data; filled markers are GPT simulations.}

\label{fig:data}
\end{figure*}
\begin{figure}
\includegraphics[width=0.95\linewidth]{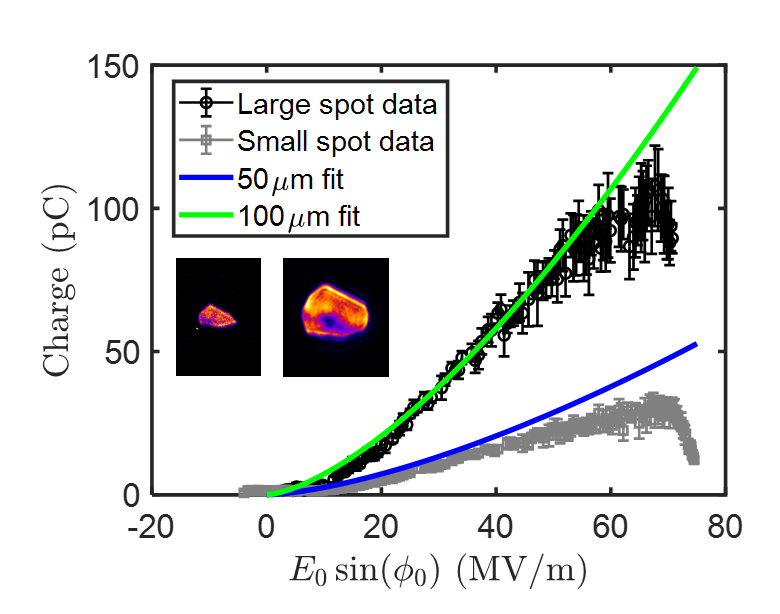}
\caption{Launch-field scan measurements for two laser spot sizes, compared with theoretical fits. Insets show the corresponding VCC profiles used to determine the effective spot radii.}
    \label{fig:spots}
\end{figure}

In order to highlight the differences in space charge saturation dynamics, we acquire phase scans at multiple laser pulse energies in both short-pulse (pancake) and long-pulse (cigar) formats. In Fig.~\ref{fig:data}(a)–(c), we show the data for three representative laser pulse energies, plotted versus the cathode field $E = E_0 \sin\phi$ with GPT simulations overlaid across launch fields from 10 to 70~MV/m. In all panels, the red curves and markers correspond to the pancake regime and the blue curves and markers correspond to the cigar regime. The VCC profile used to seed the corresponding GPT simulations is shown as an inset in Fig.~\ref{fig:data}(a).

For reference, the pancake emission is found to reach saturation at a laser energy of approximately 0.1~nJ for an extraction field of 60~MV/m. The lowest-energy scan for the pancake case shown in Fig.~\ref{fig:data}(a) (0.5~nJ) therefore already lies well above this threshold and is fully saturated across the entire field range. In contrast, at the same laser energy the cigar scan remains only partially saturated: the charge follows the space-charge–limited scaling up to approximately 40~MV/m before kinking into a QE-dominated regime. The continued rise beyond this point is attributed to Schottky-enhanced QE at higher launch fields.

Figure~\ref{fig:data}(b) shows the intermediate case (1~nJ). Here, the pancake data does not yield twice the charge of the 0.5~nJ case, confirming that emission is already well within the saturation regime. In contrast, the cigar data now follows the $E^{3/2}$ scaling predicted by Eq.~(\ref{eq:cigar}) across the measured field range, with the laser energy sufficient to reach space-charge saturation up to about 70~MV/m.

At higher laser energy (5~nJ), Fig.~\ref{fig:data}(c) shows that emission in both cases is well beyond saturation. The pancake data remains nearly linear with field, while the cigar trace begins to rise above the theoretical saturation curve of panel (b). This departure indicates that, once the cathode core is fully saturated, additional charge can still be extracted from the lower-intensity regions of the laser profile where the local field remains only partially screened. To examine this behavior more directly and to quantify how emission evolves from the QE-limited to the over-saturated regime, we next analyze the charge as a function of laser energy at a fixed extraction field.
In Figure~\ref{fig:data}(d) we replot the data showing the measured emitted charge versus laser energy at a fixed field slice around $E=60~\mathrm{MV/m}$, with blue circles for the cigar regime and red squares for the pancake regime (including error bars derived from shot-to-shot statistics) and GPT simulation results overlaid. The curves are better understood following them from left to right as we progress through the different emission regimes. At low energies, in both the pancake and cigar regime, the extracted charge increases linearly, consistent with QE-limited emission as given by Eq.~\eqref{eq:qe}. As the laser energy increases further, the data exhibits a distinct kink that marks the onset of saturation.

To connect the measured spot to the uniform-disk formulas, we introduce an effective radius $R=\sqrt{2}\,\sigma$, where $\sigma=\sqrt{\sigma_x\sigma_y}$ is the geometric mean of the rms widths in $x$ and $y$, ensuring that the effective area is preserved ($A_{\rm eff}=\pi R^2$). With $R\!\approx 25~\mu\mathrm{m}$, the saturation thresholds evaluate to $Q_{\rm sat}\!\approx 1~\mathrm{pC}$ in the pancake regime (Eq.~\eqref{eq:pancake}) and $\approx 5~\mathrm{pC}$ in the cigar regime (Eq.~\eqref{eq:cigar}). These values, shown as horizontal dashed lines in Fig.~\ref{fig:data}(d), are in good agreement with the transition points observed in the data. More exact profile-dependent formulas shift only the prefactor weakly ($\lesssim$10\% at fixed $A_{\rm eff}$), as discussed earlier in the discussion and also evident from the simulations of Fig.~\ref{fig:pulse_length_dependence}(a).

Beyond saturation, the trend changes as the process is strongly affected by the non-uniformity of the laser spot so that the central cathode region is fully saturated but the transverse tails of the laser distribution remain below threshold. In this tail-emission regime additional charge continues to be extracted not because the core emits more, but because the lower-intensity tails at the beam periphery do not fully cancel the applied field and can continue to emit more charge as the laser energy increases. To quantify this behavior we employ the tail-emission model ~\cite{ROSENZWEIG1994379},
\begin{equation}
Q_{\text{extracted}} \simeq Q_{\mathrm{sat}} \left[1 + \log\!\left(\frac{Q_{\text{expected}}}{Q_{\mathrm{sat}}}\right)\right],
\label{eq:tails}
\end{equation}
where $Q_{\text{expected}}$ is the charge predicted from QE alone. The model predicts a slow logarithmic increase beyond $Q_{\rm sat}$ for distributions with long Gaussian tails. In the pancake regime, by contrast, sharply truncated profiles (uniform or quadratic) cause the extracted charge to asymptotically approach a finite bound, for instance in the case of a perfectly uniform emission profile, $Q_{\rm sat}=\pi R^2\epsilon_0E$. In the cigar regime, however, even truncated profiles do not exhibit such a simple description of the asymptote analytically, since the peripheral regions of the cathode plane do not saturate as abruptly as the central core, as can be seen in Fig.~\ref{fig:cathodefield}(a). These behaviors are naturally included in the GPT simulation provided the actual non-uniform transverse distribution is used to initialize the beam at the cathode.
The saturation values and the tail-emission model together are used to construct a piecewise function that captures the full trend: below saturation the data follow the QE-limited expectation from Eq.~\eqref{eq:qe} (dashed purple), while beyond saturation they follow Eq.~\eqref{eq:tails}, with separate solid curves for pancake (red) and cigar (black). GPT simulations, seeded with the measured VCC profiles and shown as solid blue circle (cigar) and solid red square (pancake) markers, follow the same progression continuously through the different regimes.

Together, Eq.~\eqref{eq:pancake}, Eq.~\eqref{eq:cigar}, and Eq.~\eqref{eq:tails} provide a consistent description of the evolution from QE-dominated emission, through saturation, and into the tail-emission regime in both pancake and cigar geometries. These laser-energy scans offer a practical means to map emission profiles and identify saturation thresholds and in particular show that the cigar regime distributes the extracted charge more efficiently, enabling higher charge extraction for a given input laser energy, offering the potential for improved beam quality.
  



To test the scaling with emission area, we repeat the scans with two spots, $R\!\approx\!50~\mu\mathrm{m}$ and $100~\mu\mathrm{m}$, as determined from the VCC distributions shown in the inset. The spot sizes are set by a motorized iris imaged onto the cathode. Changing the aperture adjusts both the transmitted (post-iris) energy and the on-cathode spot size, such that the peak fluence at the cathode remains approximately constant.  
 Figure~\ref{fig:spots} shows the charge–field curves for both spots. With peak fluence high enough to drive saturation across the phase range, the extracted $Q_{\rm sat}$ is observed to scale with emission radius as $R^{3/2}$, with the extracted charge increasing by nearly a factor of $2^{3/2}$, in agreement with theory.  

From our earlier analysis, the two spot sizes tested here remain within the cigar regime. Were the iris opened further, one would eventually enter the intermediate regime between cigar and pancake aspect ratios, where the scaling deviates from what is anticipated from Eq. \ref{eq:cigar}. In performing this comparison, we avoided driving the system far beyond saturation, so as not to obscure the scaling. 
Although some extended emission may still be present, the observed agreement with the expected scaling confirms the validity of this final check and aligns with Eq. (\ref{eq:cigar}).  


\section{Conclusion}

In this work we presented to our knowledge the first direct experimental characterization of photoemission saturation in the cigar regime, where the charge–field-radius scaling $Q_{\rm sat}\propto (R E)^{3/2}$ was directly confirmed. Systematic measurements were performed at the UCLA Pegasus photoinjector, including phase scans and launch-field/laser-energy scans across varying pulse durations and spot sizes and were used to validate the analytical predictions in agreement with \textsc{General Particle Tracer} (GPT) simulations using realistic laser distributions. We clarified the role of unsaturated emission tails which can give rise to additional emission beyond the nominal saturation threshold, but note that one would probably want to avoid this situation as certain regions of the cathodes would operate in full saturation. 

Beyond validating theory, these results establish a practical characterization procedure of space charge effects in photoemission as phase scans and field/laser energy scans provide a direct way to map the saturation surface and to select operating conditions relative to it. Looking forward, a critical open question is the emittance evolution as extraction approaches saturation. Recent work shows that emittance compensation can still recover performance near the thermal limit\cite{PhysRevAccelBeams.22.023403} even close to saturation, but a predictive framework, ideally with general scaling across charge, beam aspect ratio, and field parameters is not really available. Establishing such models is especially important as the field advances toward state-of-the-art RF guns with gradients heading toward the GV/m level\cite{PhysRevAccelBeams.19.053401,PhysRevAccelBeams.22.023403,PhysRevAccelBeams.25.083402}, where understanding both the current density and emittance limits will be essential to defining the maximum achievable beam brightness in next-generation injectors.

In practice, according to simulations, operating at a fraction (e.g., a factor of two) below saturation is safe to avoid the onset of virtual-cathode instabilities and the corresponding brightness degradation. Our results, in combination with this prescription, offer then an experimentally grounded method for tuning RF photoinjectors in ultrafast electron diffraction, microscopy, and high-brightness light source applications.

\begin{acknowledgments}
This research is supported by DOE grant No. DE-SC0009914 and DE-SC0017102. We thank Theo Vecchione and Gowri Adhikari from SLAC for providing the high QE CsTe2 photocathodes without which this research would not have been possible. PD is supported by the National Science Foundation under Grant No. DMR-1548924. DG and BS are supported by the National Science Foundation under Grant No. PHY-1549132.
\end{acknowledgments}
\bibliographystyle{apsrev4-2}
\bibliography{references_p}

\begin{thebibliography}{38}%
\makeatletter
\providecommand \@ifxundefined [1]{%
 \@ifx{#1\undefined}
}%
\providecommand \@ifnum [1]{%
 \ifnum #1\expandafter \@firstoftwo
 \else \expandafter \@secondoftwo
 \fi
}%
\providecommand \@ifx [1]{%
 \ifx #1\expandafter \@firstoftwo
 \else \expandafter \@secondoftwo
 \fi
}%
\providecommand \natexlab [1]{#1}%
\providecommand \enquote  [1]{``#1''}%
\providecommand \bibnamefont  [1]{#1}%
\providecommand \bibfnamefont [1]{#1}%
\providecommand \citenamefont [1]{#1}%
\providecommand \href@noop [0]{\@secondoftwo}%
\providecommand \href [0]{\begingroup \@sanitize@url \@href}%
\providecommand \@href[1]{\@@startlink{#1}\@@href}%
\providecommand \@@href[1]{\endgroup#1\@@endlink}%
\providecommand \@sanitize@url [0]{\catcode `\\12\catcode `\$12\catcode `\&12\catcode `\#12\catcode `\^12\catcode `\_12\catcode `\%12\relax}%
\providecommand \@@startlink[1]{}%
\providecommand \@@endlink[0]{}%
\providecommand \url  [0]{\begingroup\@sanitize@url \@url }%
\providecommand \@url [1]{\endgroup\@href {#1}{\urlprefix }}%
\providecommand \urlprefix  [0]{URL }%
\providecommand \Eprint [0]{\href }%
\providecommand \doibase [0]{https://doi.org/}%
\providecommand \selectlanguage [0]{\@gobble}%
\providecommand \bibinfo  [0]{\@secondoftwo}%
\providecommand \bibfield  [0]{\@secondoftwo}%
\providecommand \translation [1]{[#1]}%
\providecommand \BibitemOpen [0]{}%
\providecommand \bibitemStop [0]{}%
\providecommand \bibitemNoStop [0]{.\EOS\space}%
\providecommand \EOS [0]{\spacefactor3000\relax}%
\providecommand \BibitemShut  [1]{\csname bibitem#1\endcsname}%
\let\auto@bib@innerbib\@empty
\bibitem [{\citenamefont {Bahrdt}\ \emph {et~al.}(2013)\citenamefont {Bahrdt}, \citenamefont {Holldack}, \citenamefont {Kuske}, \citenamefont {M{\"{u}}ller}, \citenamefont {Scheer},\ and\ \citenamefont {Schmid}}]{Bahrdt2013FirstRadiation}%
  \BibitemOpen
  \bibfield  {author} {\bibinfo {author} {\bibfnamefont {J.}~\bibnamefont {Bahrdt}}, \bibinfo {author} {\bibfnamefont {K.}~\bibnamefont {Holldack}}, \bibinfo {author} {\bibfnamefont {P.}~\bibnamefont {Kuske}}, \bibinfo {author} {\bibfnamefont {R.}~\bibnamefont {M{\"{u}}ller}}, \bibinfo {author} {\bibfnamefont {M.}~\bibnamefont {Scheer}},\ and\ \bibinfo {author} {\bibfnamefont {P.}~\bibnamefont {Schmid}},\ }\href {https://doi.org/10.1103/PHYSREVLETT.111.034801/FIGURES/6/MEDIUM} {\bibfield  {journal} {\bibinfo  {journal} {Physical Review Letters}\ }\textbf {\bibinfo {volume} {111}},\ \bibinfo {pages} {034801} (\bibinfo {year} {2013})}\BibitemShut {NoStop}%
\bibitem [{\citenamefont {Jentschura}\ and\ \citenamefont {Serbo}(2011)}]{Jentschura2011GenerationBackscattering}%
  \BibitemOpen
  \bibfield  {author} {\bibinfo {author} {\bibfnamefont {U.~D.}\ \bibnamefont {Jentschura}}\ and\ \bibinfo {author} {\bibfnamefont {V.~G.}\ \bibnamefont {Serbo}},\ }\href {https://doi.org/10.1103/PHYSREVLETT.106.013001/FIGURES/2/MEDIUM} {\bibfield  {journal} {\bibinfo  {journal} {Physical Review Letters}\ }\textbf {\bibinfo {volume} {106}},\ \bibinfo {pages} {013001} (\bibinfo {year} {2011})}\BibitemShut {NoStop}%
\bibitem [{\citenamefont {Shen}\ \emph {et~al.}(2011)\citenamefont {Shen}, \citenamefont {Yang}, \citenamefont {Carr}, \citenamefont {Hidaka}, \citenamefont {Murphy},\ and\ \citenamefont {Wang}}]{PhysRevLett.107.204801}%
  \BibitemOpen
  \bibfield  {author} {\bibinfo {author} {\bibfnamefont {Y.}~\bibnamefont {Shen}}, \bibinfo {author} {\bibfnamefont {X.}~\bibnamefont {Yang}}, \bibinfo {author} {\bibfnamefont {G.~L.}\ \bibnamefont {Carr}}, \bibinfo {author} {\bibfnamefont {Y.}~\bibnamefont {Hidaka}}, \bibinfo {author} {\bibfnamefont {J.~B.}\ \bibnamefont {Murphy}},\ and\ \bibinfo {author} {\bibfnamefont {X.}~\bibnamefont {Wang}},\ }\href {https://doi.org/10.1103/PhysRevLett.107.204801} {\bibfield  {journal} {\bibinfo  {journal} {Phys. Rev. Lett.}\ }\textbf {\bibinfo {volume} {107}},\ \bibinfo {pages} {204801} (\bibinfo {year} {2011})}\BibitemShut {NoStop}%
\bibitem [{\citenamefont {Filippetto}\ \emph {et~al.}(2022)\citenamefont {Filippetto}, \citenamefont {Musumeci}, \citenamefont {Li}, \citenamefont {Siwick}, \citenamefont {Otto}, \citenamefont {Centurion},\ and\ \citenamefont {Nunes}}]{RevModPhys.94.045004}%
  \BibitemOpen
  \bibfield  {author} {\bibinfo {author} {\bibfnamefont {D.}~\bibnamefont {Filippetto}}, \bibinfo {author} {\bibfnamefont {P.}~\bibnamefont {Musumeci}}, \bibinfo {author} {\bibfnamefont {R.~K.}\ \bibnamefont {Li}}, \bibinfo {author} {\bibfnamefont {B.~J.}\ \bibnamefont {Siwick}}, \bibinfo {author} {\bibfnamefont {M.~R.}\ \bibnamefont {Otto}}, \bibinfo {author} {\bibfnamefont {M.}~\bibnamefont {Centurion}},\ and\ \bibinfo {author} {\bibfnamefont {J.~P.~F.}\ \bibnamefont {Nunes}},\ }\href {https://doi.org/10.1103/RevModPhys.94.045004} {\bibfield  {journal} {\bibinfo  {journal} {Rev. Mod. Phys.}\ }\textbf {\bibinfo {volume} {94}},\ \bibinfo {pages} {045004} (\bibinfo {year} {2022})}\BibitemShut {NoStop}%
\bibitem [{\citenamefont {Wu}\ \emph {et~al.}(2021)\citenamefont {Wu}, \citenamefont {Hua}, \citenamefont {Zhou}, \citenamefont {Zhang}, \citenamefont {Liu}, \citenamefont {Peng}, \citenamefont {Fang}, \citenamefont {Ning}, \citenamefont {Nie}, \citenamefont {Li} \emph {et~al.}}]{wu2021high}%
  \BibitemOpen
  \bibfield  {author} {\bibinfo {author} {\bibfnamefont {Y.}~\bibnamefont {Wu}}, \bibinfo {author} {\bibfnamefont {J.}~\bibnamefont {Hua}}, \bibinfo {author} {\bibfnamefont {Z.}~\bibnamefont {Zhou}}, \bibinfo {author} {\bibfnamefont {J.}~\bibnamefont {Zhang}}, \bibinfo {author} {\bibfnamefont {S.}~\bibnamefont {Liu}}, \bibinfo {author} {\bibfnamefont {B.}~\bibnamefont {Peng}}, \bibinfo {author} {\bibfnamefont {Y.}~\bibnamefont {Fang}}, \bibinfo {author} {\bibfnamefont {X.}~\bibnamefont {Ning}}, \bibinfo {author} {\bibfnamefont {Z.}~\bibnamefont {Nie}}, \bibinfo {author} {\bibfnamefont {F.}~\bibnamefont {Li}}, \emph {et~al.},\ }\href@noop {} {\bibfield  {journal} {\bibinfo  {journal} {Nature Physics}\ }\textbf {\bibinfo {volume} {17}},\ \bibinfo {pages} {801} (\bibinfo {year} {2021})}\BibitemShut {NoStop}%
\bibitem [{\citenamefont {Marsh}\ \emph {et~al.}(2018)\citenamefont {Marsh}, \citenamefont {Anderson}, \citenamefont {Anderson}, \citenamefont {Gibson}, \citenamefont {Barty},\ and\ \citenamefont {Hwang}}]{PhysRevAccelBeams.21.073401}%
  \BibitemOpen
  \bibfield  {author} {\bibinfo {author} {\bibfnamefont {R.~A.}\ \bibnamefont {Marsh}}, \bibinfo {author} {\bibfnamefont {G.~G.}\ \bibnamefont {Anderson}}, \bibinfo {author} {\bibfnamefont {S.~G.}\ \bibnamefont {Anderson}}, \bibinfo {author} {\bibfnamefont {D.~J.}\ \bibnamefont {Gibson}}, \bibinfo {author} {\bibfnamefont {C.~P.~J.}\ \bibnamefont {Barty}},\ and\ \bibinfo {author} {\bibfnamefont {Y.}~\bibnamefont {Hwang}},\ }\href {https://doi.org/10.1103/PhysRevAccelBeams.21.073401} {\bibfield  {journal} {\bibinfo  {journal} {Phys. Rev. Accel. Beams}\ }\textbf {\bibinfo {volume} {21}},\ \bibinfo {pages} {073401} (\bibinfo {year} {2018})}\BibitemShut {NoStop}%
\bibitem [{\citenamefont {Ha}\ \emph {et~al.}(2022)\citenamefont {Ha}, \citenamefont {Kim}, \citenamefont {Power}, \citenamefont {Sun},\ and\ \citenamefont {Piot}}]{RevModPhys.94.025006}%
  \BibitemOpen
  \bibfield  {author} {\bibinfo {author} {\bibfnamefont {G.}~\bibnamefont {Ha}}, \bibinfo {author} {\bibfnamefont {K.-J.}\ \bibnamefont {Kim}}, \bibinfo {author} {\bibfnamefont {J.~G.}\ \bibnamefont {Power}}, \bibinfo {author} {\bibfnamefont {Y.}~\bibnamefont {Sun}},\ and\ \bibinfo {author} {\bibfnamefont {P.}~\bibnamefont {Piot}},\ }\href {https://doi.org/10.1103/RevModPhys.94.025006} {\bibfield  {journal} {\bibinfo  {journal} {Rev. Mod. Phys.}\ }\textbf {\bibinfo {volume} {94}},\ \bibinfo {pages} {025006} (\bibinfo {year} {2022})}\BibitemShut {NoStop}%
\bibitem [{\citenamefont {Hemsing}\ \emph {et~al.}(2014)\citenamefont {Hemsing}, \citenamefont {Stupakov}, \citenamefont {Xiang},\ and\ \citenamefont {Zholents}}]{RevModPhys.86.897}%
  \BibitemOpen
  \bibfield  {author} {\bibinfo {author} {\bibfnamefont {E.}~\bibnamefont {Hemsing}}, \bibinfo {author} {\bibfnamefont {G.}~\bibnamefont {Stupakov}}, \bibinfo {author} {\bibfnamefont {D.}~\bibnamefont {Xiang}},\ and\ \bibinfo {author} {\bibfnamefont {A.}~\bibnamefont {Zholents}},\ }\href {https://doi.org/10.1103/RevModPhys.86.897} {\bibfield  {journal} {\bibinfo  {journal} {Rev. Mod. Phys.}\ }\textbf {\bibinfo {volume} {86}},\ \bibinfo {pages} {897} (\bibinfo {year} {2014})}\BibitemShut {NoStop}%
\bibitem [{\citenamefont {Musumeci}\ \emph {et~al.}(2018)\citenamefont {Musumeci}, \citenamefont {Navarro}, \citenamefont {Rosenzweig}, \citenamefont {Cultrera}, \citenamefont {Bazarov}, \citenamefont {Maxson}, \citenamefont {Karkare},\ and\ \citenamefont {Padmore}}]{musumeci2018advances}%
  \BibitemOpen
  \bibfield  {author} {\bibinfo {author} {\bibfnamefont {P.}~\bibnamefont {Musumeci}}, \bibinfo {author} {\bibfnamefont {J.~G.}\ \bibnamefont {Navarro}}, \bibinfo {author} {\bibfnamefont {J.}~\bibnamefont {Rosenzweig}}, \bibinfo {author} {\bibfnamefont {L.}~\bibnamefont {Cultrera}}, \bibinfo {author} {\bibfnamefont {I.}~\bibnamefont {Bazarov}}, \bibinfo {author} {\bibfnamefont {J.}~\bibnamefont {Maxson}}, \bibinfo {author} {\bibfnamefont {S.}~\bibnamefont {Karkare}},\ and\ \bibinfo {author} {\bibfnamefont {H.}~\bibnamefont {Padmore}},\ }\href@noop {} {\bibfield  {journal} {\bibinfo  {journal} {Nuclear Instruments and Methods in Physics Research Section A: Accelerators, Spectrometers, Detectors and Associated Equipment}\ }\textbf {\bibinfo {volume} {907}},\ \bibinfo {pages} {209} (\bibinfo {year} {2018})}\BibitemShut {NoStop}%
\bibitem [{\citenamefont {Bazarov}\ \emph {et~al.}(2009)\citenamefont {Bazarov}, \citenamefont {Dunham},\ and\ \citenamefont {Sinclair}}]{PhysRevLett.102.104801}%
  \BibitemOpen
  \bibfield  {author} {\bibinfo {author} {\bibfnamefont {I.~V.}\ \bibnamefont {Bazarov}}, \bibinfo {author} {\bibfnamefont {B.~M.}\ \bibnamefont {Dunham}},\ and\ \bibinfo {author} {\bibfnamefont {C.~K.}\ \bibnamefont {Sinclair}},\ }\href {https://doi.org/10.1103/PhysRevLett.102.104801} {\bibfield  {journal} {\bibinfo  {journal} {Phys. Rev. Lett.}\ }\textbf {\bibinfo {volume} {102}},\ \bibinfo {pages} {104801} (\bibinfo {year} {2009})}\BibitemShut {NoStop}%
\bibitem [{\citenamefont {Filippetto}\ \emph {et~al.}(2014)\citenamefont {Filippetto}, \citenamefont {Musumeci}, \citenamefont {Zolotorev},\ and\ \citenamefont {Stupakov}}]{PhysRevSTAB.17.024201}%
  \BibitemOpen
  \bibfield  {author} {\bibinfo {author} {\bibfnamefont {D.}~\bibnamefont {Filippetto}}, \bibinfo {author} {\bibfnamefont {P.}~\bibnamefont {Musumeci}}, \bibinfo {author} {\bibfnamefont {M.}~\bibnamefont {Zolotorev}},\ and\ \bibinfo {author} {\bibfnamefont {G.}~\bibnamefont {Stupakov}},\ }\href {https://doi.org/10.1103/PhysRevSTAB.17.024201} {\bibfield  {journal} {\bibinfo  {journal} {Phys. Rev. ST Accel. Beams}\ }\textbf {\bibinfo {volume} {17}},\ \bibinfo {pages} {024201} (\bibinfo {year} {2014})}\BibitemShut {NoStop}%
\bibitem [{\citenamefont {Child}(1911)}]{PhysRevSeriesI.32.492}%
  \BibitemOpen
  \bibfield  {author} {\bibinfo {author} {\bibfnamefont {C.~D.}\ \bibnamefont {Child}},\ }\href {https://doi.org/10.1103/PhysRevSeriesI.32.492} {\bibfield  {journal} {\bibinfo  {journal} {Phys. Rev. (Series I)}\ }\textbf {\bibinfo {volume} {32}},\ \bibinfo {pages} {492} (\bibinfo {year} {1911})}\BibitemShut {NoStop}%
\bibitem [{\citenamefont {Zhang}\ \emph {et~al.}(2021)\citenamefont {Zhang}, \citenamefont {Ang}, \citenamefont {Garner}, \citenamefont {Valfells}, \citenamefont {Luginsland},\ and\ \citenamefont {Ang}}]{10.1063/5.0042355}%
  \BibitemOpen
  \bibfield  {author} {\bibinfo {author} {\bibfnamefont {P.}~\bibnamefont {Zhang}}, \bibinfo {author} {\bibfnamefont {Y.~S.}\ \bibnamefont {Ang}}, \bibinfo {author} {\bibfnamefont {A.~L.}\ \bibnamefont {Garner}}, \bibinfo {author} {\bibfnamefont {A.}~\bibnamefont {Valfells}}, \bibinfo {author} {\bibfnamefont {J.~W.}\ \bibnamefont {Luginsland}},\ and\ \bibinfo {author} {\bibfnamefont {L.~K.}\ \bibnamefont {Ang}},\ }\href {https://doi.org/10.1063/5.0042355} {\bibfield  {journal} {\bibinfo  {journal} {Journal of Applied Physics}\ }\textbf {\bibinfo {volume} {129}},\ \bibinfo {pages} {100902} (\bibinfo {year} {2021})}\BibitemShut {NoStop}%
\bibitem [{\citenamefont {Langmuir}(1923)}]{PhysRev.21.419}%
  \BibitemOpen
  \bibfield  {author} {\bibinfo {author} {\bibfnamefont {I.}~\bibnamefont {Langmuir}},\ }\href {https://doi.org/10.1103/PhysRev.21.419} {\bibfield  {journal} {\bibinfo  {journal} {Phys. Rev.}\ }\textbf {\bibinfo {volume} {21}},\ \bibinfo {pages} {419} (\bibinfo {year} {1923})}\BibitemShut {NoStop}%
\bibitem [{\citenamefont {Griswold}\ and\ \citenamefont {Fisch}(2016)}]{10.1063/1.4939607}%
  \BibitemOpen
  \bibfield  {author} {\bibinfo {author} {\bibfnamefont {M.~E.}\ \bibnamefont {Griswold}}\ and\ \bibinfo {author} {\bibfnamefont {N.~J.}\ \bibnamefont {Fisch}},\ }\href {https://doi.org/10.1063/1.4939607} {\bibfield  {journal} {\bibinfo  {journal} {Physics of Plasmas}\ }\textbf {\bibinfo {volume} {23}},\ \bibinfo {pages} {014502} (\bibinfo {year} {2016})},\ \Eprint {https://arxiv.org/abs/https://pubs.aip.org/aip/pop/article-pdf/doi/10.1063/1.4939607/15718060/014502\_1\_online.pdf} {https://pubs.aip.org/aip/pop/article-pdf/doi/10.1063/1.4939607/15718060/014502\_1\_online.pdf} \BibitemShut {NoStop}%
\bibitem [{\citenamefont {Caflisch}\ and\ \citenamefont {Rosin}(2012)}]{PhysRevE.85.056408}%
  \BibitemOpen
  \bibfield  {author} {\bibinfo {author} {\bibfnamefont {R.~E.}\ \bibnamefont {Caflisch}}\ and\ \bibinfo {author} {\bibfnamefont {M.~S.}\ \bibnamefont {Rosin}},\ }\href {https://doi.org/10.1103/PhysRevE.85.056408} {\bibfield  {journal} {\bibinfo  {journal} {Phys. Rev. E}\ }\textbf {\bibinfo {volume} {85}},\ \bibinfo {pages} {056408} (\bibinfo {year} {2012})}\BibitemShut {NoStop}%
\bibitem [{\citenamefont {Valfells}\ \emph {et~al.}(2002)\citenamefont {Valfells}, \citenamefont {Feldman}, \citenamefont {Virgo}, \citenamefont {O’Shea},\ and\ \citenamefont {Lau}}]{10.1063/1.1463065}%
  \BibitemOpen
  \bibfield  {author} {\bibinfo {author} {\bibfnamefont {A.}~\bibnamefont {Valfells}}, \bibinfo {author} {\bibfnamefont {D.~W.}\ \bibnamefont {Feldman}}, \bibinfo {author} {\bibfnamefont {M.}~\bibnamefont {Virgo}}, \bibinfo {author} {\bibfnamefont {P.~G.}\ \bibnamefont {O’Shea}},\ and\ \bibinfo {author} {\bibfnamefont {Y.~Y.}\ \bibnamefont {Lau}},\ }\href {https://doi.org/10.1063/1.1463065} {\bibfield  {journal} {\bibinfo  {journal} {Physics of Plasmas}\ }\textbf {\bibinfo {volume} {9}},\ \bibinfo {pages} {2377} (\bibinfo {year} {2002})}\BibitemShut {NoStop}%
\bibitem [{\citenamefont {Gunnarsson}\ \emph {et~al.}(2020)\citenamefont {Gunnarsson}, \citenamefont {Torfason}, \citenamefont {Manolescu},\ and\ \citenamefont {Valfells}}]{Gunnarsson:2020bmf}%
  \BibitemOpen
  \bibfield  {author} {\bibinfo {author} {\bibfnamefont {J.~B.}\ \bibnamefont {Gunnarsson}}, \bibinfo {author} {\bibfnamefont {K.}~\bibnamefont {Torfason}}, \bibinfo {author} {\bibfnamefont {A.}~\bibnamefont {Manolescu}},\ and\ \bibinfo {author} {\bibfnamefont {A.}~\bibnamefont {Valfells}},\ }\href {https://doi.org/10.1109/TED.2020.3037280} {\bibfield  {journal} {\bibinfo  {journal} {IEEE Trans. Electron. Dev.}\ }\textbf {\bibinfo {volume} {68}},\ \bibinfo {pages} {342} (\bibinfo {year} {2020})},\ \Eprint {https://arxiv.org/abs/2010.01334} {arXiv:2010.01334 [physics.acc-ph]} \BibitemShut {NoStop}%
\bibitem [{\citenamefont {Schneider}\ \emph {et~al.}(2021)\citenamefont {Schneider}, \citenamefont {Sims}, \citenamefont {Jevarjian}, \citenamefont {Shinohara}, \citenamefont {Nikhar}, \citenamefont {Posos}, \citenamefont {Liu}, \citenamefont {Power}, \citenamefont {Shao},\ and\ \citenamefont {Baryshev}}]{PhysRevAccelBeams.24.123401}%
  \BibitemOpen
  \bibfield  {author} {\bibinfo {author} {\bibfnamefont {M.~E.}\ \bibnamefont {Schneider}}, \bibinfo {author} {\bibfnamefont {B.}~\bibnamefont {Sims}}, \bibinfo {author} {\bibfnamefont {E.}~\bibnamefont {Jevarjian}}, \bibinfo {author} {\bibfnamefont {R.}~\bibnamefont {Shinohara}}, \bibinfo {author} {\bibfnamefont {T.}~\bibnamefont {Nikhar}}, \bibinfo {author} {\bibfnamefont {T.~Y.}\ \bibnamefont {Posos}}, \bibinfo {author} {\bibfnamefont {W.}~\bibnamefont {Liu}}, \bibinfo {author} {\bibfnamefont {J.}~\bibnamefont {Power}}, \bibinfo {author} {\bibfnamefont {J.}~\bibnamefont {Shao}},\ and\ \bibinfo {author} {\bibfnamefont {S.~V.}\ \bibnamefont {Baryshev}},\ }\href {https://doi.org/10.1103/PhysRevAccelBeams.24.123401} {\bibfield  {journal} {\bibinfo  {journal} {Phys. Rev. Accel. Beams}\ }\textbf {\bibinfo {volume} {24}},\ \bibinfo {pages} {123401} (\bibinfo {year} {2021})}\BibitemShut {NoStop}%
\bibitem [{gpt()}]{gpt}%
  \BibitemOpen
  \href@noop {} {\bibinfo {title} {{General Particle Tracer}}},\ \bibinfo {howpublished} {\url{https://www.pulsar.nl/gpt/index.html}}\BibitemShut {NoStop}%
\bibitem [{\citenamefont {Coutsias}\ and\ \citenamefont {Sullivan}(1983)}]{PhysRevA.27.1535}%
  \BibitemOpen
  \bibfield  {author} {\bibinfo {author} {\bibfnamefont {E.~A.}\ \bibnamefont {Coutsias}}\ and\ \bibinfo {author} {\bibfnamefont {D.~J.}\ \bibnamefont {Sullivan}},\ }\href {https://doi.org/10.1103/PhysRevA.27.1535} {\bibfield  {journal} {\bibinfo  {journal} {Phys. Rev. A}\ }\textbf {\bibinfo {volume} {27}},\ \bibinfo {pages} {1535} (\bibinfo {year} {1983})}\BibitemShut {NoStop}%
\bibitem [{\citenamefont {Birdsall}\ and\ \citenamefont {Bridges}(1961)}]{10.1063/1.1728361}%
  \BibitemOpen
  \bibfield  {author} {\bibinfo {author} {\bibfnamefont {C.~K.}\ \bibnamefont {Birdsall}}\ and\ \bibinfo {author} {\bibfnamefont {W.~B.}\ \bibnamefont {Bridges}},\ }\href {https://doi.org/10.1063/1.1728361} {\bibfield  {journal} {\bibinfo  {journal} {Journal of Applied Physics}\ }\textbf {\bibinfo {volume} {32}},\ \bibinfo {pages} {2611} (\bibinfo {year} {1961})}\BibitemShut {NoStop}%
\bibitem [{\citenamefont {Jackson}(1998)}]{jackson3}%
  \BibitemOpen
  \bibfield  {author} {\bibinfo {author} {\bibfnamefont {J.~D.}\ \bibnamefont {Jackson}},\ }\href {https://doi.org/https://doi.org/10.1002/3527600434.eap109} {\emph {\bibinfo {title} {{Classical Electrodynamics, Third Edition}}}}\ (\bibinfo {year} {1998})\BibitemShut {NoStop}%
\bibitem [{\citenamefont {Shamuilov}\ \emph {et~al.}(2018)\citenamefont {Shamuilov}, \citenamefont {Mak}, \citenamefont {Pepitone},\ and\ \citenamefont {Goryashko}}]{10.1063/1.5063888}%
  \BibitemOpen
  \bibfield  {author} {\bibinfo {author} {\bibfnamefont {G.}~\bibnamefont {Shamuilov}}, \bibinfo {author} {\bibfnamefont {A.}~\bibnamefont {Mak}}, \bibinfo {author} {\bibfnamefont {K.}~\bibnamefont {Pepitone}},\ and\ \bibinfo {author} {\bibfnamefont {V.}~\bibnamefont {Goryashko}},\ }\href {https://doi.org/10.1063/1.5063888} {\bibfield  {journal} {\bibinfo  {journal} {Applied Physics Letters}\ }\textbf {\bibinfo {volume} {113}},\ \bibinfo {pages} {204103} (\bibinfo {year} {2018})},\ \Eprint {https://arxiv.org/abs/https://pubs.aip.org/aip/apl/article-pdf/doi/10.1063/1.5063888/14520529/204103\_1\_online.pdf} {https://pubs.aip.org/aip/apl/article-pdf/doi/10.1063/1.5063888/14520529/204103\_1\_online.pdf} \BibitemShut {NoStop}%
\bibitem [{\citenamefont {Lau}(2001)}]{PhysRevLett.87.278301}%
  \BibitemOpen
  \bibfield  {author} {\bibinfo {author} {\bibfnamefont {Y.~Y.}\ \bibnamefont {Lau}},\ }\href {https://doi.org/10.1103/PhysRevLett.87.278301} {\bibfield  {journal} {\bibinfo  {journal} {Phys. Rev. Lett.}\ }\textbf {\bibinfo {volume} {87}},\ \bibinfo {pages} {278301} (\bibinfo {year} {2001})}\BibitemShut {NoStop}%
\bibitem [{\citenamefont {Stulle}\ and\ \citenamefont {Bergoz}(2015)}]{stulle2015turbo}%
  \BibitemOpen
  \bibfield  {author} {\bibinfo {author} {\bibfnamefont {F.}~\bibnamefont {Stulle}}\ and\ \bibinfo {author} {\bibfnamefont {J.}~\bibnamefont {Bergoz}},\ }in\ \href {https://accelconf.web.cern.ch/FEL2015/papers/mop041.pdf} {\emph {\bibinfo {booktitle} {Proceedings of FEL2015}}}\ (\bibinfo {organization} {Bergoz Instrumentation, Saint-Genis-Pouilly, France},\ \bibinfo {year} {2015})\ p.\ \bibinfo {pages} {MOP041}\BibitemShut {NoStop}%
\bibitem [{\citenamefont {Alesini}\ \emph {et~al.}(2015)\citenamefont {Alesini}, \citenamefont {Battisti}, \citenamefont {Ferrario}, \citenamefont {Foggetta}, \citenamefont {Lollo}, \citenamefont {Ficcadenti}, \citenamefont {Pettinacci}, \citenamefont {Custodio}, \citenamefont {Pirez}, \citenamefont {Musumeci},\ and\ \citenamefont {Palumbo}}]{alesini:pegasus}%
  \BibitemOpen
  \bibfield  {author} {\bibinfo {author} {\bibfnamefont {D.}~\bibnamefont {Alesini}}, \bibinfo {author} {\bibfnamefont {A.}~\bibnamefont {Battisti}}, \bibinfo {author} {\bibfnamefont {M.}~\bibnamefont {Ferrario}}, \bibinfo {author} {\bibfnamefont {L.}~\bibnamefont {Foggetta}}, \bibinfo {author} {\bibfnamefont {V.}~\bibnamefont {Lollo}}, \bibinfo {author} {\bibfnamefont {L.}~\bibnamefont {Ficcadenti}}, \bibinfo {author} {\bibfnamefont {V.}~\bibnamefont {Pettinacci}}, \bibinfo {author} {\bibfnamefont {S.}~\bibnamefont {Custodio}}, \bibinfo {author} {\bibfnamefont {E.}~\bibnamefont {Pirez}}, \bibinfo {author} {\bibfnamefont {P.}~\bibnamefont {Musumeci}},\ and\ \bibinfo {author} {\bibfnamefont {L.}~\bibnamefont {Palumbo}},\ }\href {https://doi.org/10.1103/PhysRevSTAB.18.092001} {\bibfield  {journal} {\bibinfo  {journal} {Phys. Rev. ST Accel. Beams}\ }\textbf {\bibinfo {volume} {18}},\ \bibinfo {pages} {092001} (\bibinfo {year} {2015})}\BibitemShut {NoStop}%
\bibitem [{\citenamefont {Musumeci}\ \emph {et~al.}(2010)\citenamefont {Musumeci}, \citenamefont {Moody}, \citenamefont {Scoby}, \citenamefont {Gutierrez}, \citenamefont {Westfall},\ and\ \citenamefont {Li}}]{musumeci2010capturing}%
  \BibitemOpen
  \bibfield  {author} {\bibinfo {author} {\bibfnamefont {P.}~\bibnamefont {Musumeci}}, \bibinfo {author} {\bibfnamefont {J.}~\bibnamefont {Moody}}, \bibinfo {author} {\bibfnamefont {C.}~\bibnamefont {Scoby}}, \bibinfo {author} {\bibfnamefont {M.}~\bibnamefont {Gutierrez}}, \bibinfo {author} {\bibfnamefont {M.}~\bibnamefont {Westfall}},\ and\ \bibinfo {author} {\bibfnamefont {R.}~\bibnamefont {Li}},\ }\href@noop {} {\bibfield  {journal} {\bibinfo  {journal} {Journal of Applied Physics}\ }\textbf {\bibinfo {volume} {108}} (\bibinfo {year} {2010})}\BibitemShut {NoStop}%
\bibitem [{\citenamefont {Prat}\ \emph {et~al.}(2015)\citenamefont {Prat}, \citenamefont {Bettoni}, \citenamefont {Braun}, \citenamefont {Divall},\ and\ \citenamefont {Schietinger}}]{article_prat}%
  \BibitemOpen
  \bibfield  {author} {\bibinfo {author} {\bibfnamefont {E.}~\bibnamefont {Prat}}, \bibinfo {author} {\bibfnamefont {S.}~\bibnamefont {Bettoni}}, \bibinfo {author} {\bibfnamefont {H.-H.}\ \bibnamefont {Braun}}, \bibinfo {author} {\bibfnamefont {M.}~\bibnamefont {Divall}},\ and\ \bibinfo {author} {\bibfnamefont {T.}~\bibnamefont {Schietinger}},\ }\href {https://doi.org/10.1103/PhysRevSTAB.18.063401} {\bibfield  {journal} {\bibinfo  {journal} {Phys. Rev. ST Accel. Beams}\ }\textbf {\bibinfo {volume} {18}},\ \bibinfo {pages} {063401} (\bibinfo {year} {2015})}\BibitemShut {NoStop}%
\bibitem [{\citenamefont {Maxson}\ \emph {et~al.}(2017)\citenamefont {Maxson}, \citenamefont {Cesar}, \citenamefont {Calmasini}, \citenamefont {Ody}, \citenamefont {Musumeci},\ and\ \citenamefont {Alesini}}]{PhysRevLett.118.154802}%
  \BibitemOpen
  \bibfield  {author} {\bibinfo {author} {\bibfnamefont {J.}~\bibnamefont {Maxson}}, \bibinfo {author} {\bibfnamefont {D.}~\bibnamefont {Cesar}}, \bibinfo {author} {\bibfnamefont {G.}~\bibnamefont {Calmasini}}, \bibinfo {author} {\bibfnamefont {A.}~\bibnamefont {Ody}}, \bibinfo {author} {\bibfnamefont {P.}~\bibnamefont {Musumeci}},\ and\ \bibinfo {author} {\bibfnamefont {D.}~\bibnamefont {Alesini}},\ }\href {https://doi.org/10.1103/PhysRevLett.118.154802} {\bibfield  {journal} {\bibinfo  {journal} {Phys. Rev. Lett.}\ }\textbf {\bibinfo {volume} {118}},\ \bibinfo {pages} {154802} (\bibinfo {year} {2017})}\BibitemShut {NoStop}%
\bibitem [{\citenamefont {Schaap}\ and\ \citenamefont {Musumeci}(2025)}]{schaap:2025}%
  \BibitemOpen
  \bibfield  {author} {\bibinfo {author} {\bibfnamefont {B.~H.}\ \bibnamefont {Schaap}}\ and\ \bibinfo {author} {\bibfnamefont {P.}~\bibnamefont {Musumeci}},\ }\href {https://doi.org/10.1103/PhysRevAccelBeams.28.012802} {\bibfield  {journal} {\bibinfo  {journal} {Phys. Rev. Accel. Beams}\ }\textbf {\bibinfo {volume} {28}},\ \bibinfo {pages} {012802} (\bibinfo {year} {2025})}\BibitemShut {NoStop}%
\bibitem [{\citenamefont {Schmerge}\ \emph {et~al.}(2006)\citenamefont {Schmerge}, \citenamefont {Clendenin}, \citenamefont {Dowell},\ and\ \citenamefont {Gierman}}]{schmerge2006rf}%
  \BibitemOpen
  \bibfield  {author} {\bibinfo {author} {\bibfnamefont {J.~F.}\ \bibnamefont {Schmerge}}, \bibinfo {author} {\bibfnamefont {J.~E.}\ \bibnamefont {Clendenin}}, \bibinfo {author} {\bibfnamefont {D.~H.}\ \bibnamefont {Dowell}},\ and\ \bibinfo {author} {\bibfnamefont {S.~M.}\ \bibnamefont {Gierman}},\ }\href {https://www.slac.stanford.edu/pubs/slacpubs/11500/slac-pub-11700.pdf} {\emph {\bibinfo {title} {RF Gun Photo-Emission Model for Metal Cathodes Including Time Dependent Emission}}},\ \bibinfo {type} {Tech. Rep.}\ \bibinfo {number} {SLAC-PUB-11700}\ (\bibinfo  {institution} {SLAC, Stanford University},\ \bibinfo {address} {2575 Sand Hill Rd, Menlo Park, CA 94025, USA},\ \bibinfo {year} {2006})\ \bibinfo {note} {presented at the 2006 International Linear Collider Workshop (LCWS06), Bangalore, India, March 2006}\BibitemShut {NoStop}%
\bibitem [{\citenamefont {Dowell}\ and\ \citenamefont {Schmerge}(2009)}]{PhysRevSTAB.12.074201}%
  \BibitemOpen
  \bibfield  {author} {\bibinfo {author} {\bibfnamefont {D.~H.}\ \bibnamefont {Dowell}}\ and\ \bibinfo {author} {\bibfnamefont {J.~F.}\ \bibnamefont {Schmerge}},\ }\href {https://doi.org/10.1103/PhysRevSTAB.12.074201} {\bibfield  {journal} {\bibinfo  {journal} {Phys. Rev. ST Accel. Beams}\ }\textbf {\bibinfo {volume} {12}},\ \bibinfo {pages} {074201} (\bibinfo {year} {2009})}\BibitemShut {NoStop}%
\bibitem [{\citenamefont {Pierce}\ \emph {et~al.}(2021)\citenamefont {Pierce}, \citenamefont {Bae}, \citenamefont {Galdi}, \citenamefont {Cultrera}, \citenamefont {Bazarov},\ and\ \citenamefont {Maxson}}]{pierce2021beam}%
  \BibitemOpen
  \bibfield  {author} {\bibinfo {author} {\bibfnamefont {C.~M.}\ \bibnamefont {Pierce}}, \bibinfo {author} {\bibfnamefont {J.~K.}\ \bibnamefont {Bae}}, \bibinfo {author} {\bibfnamefont {A.}~\bibnamefont {Galdi}}, \bibinfo {author} {\bibfnamefont {L.}~\bibnamefont {Cultrera}}, \bibinfo {author} {\bibfnamefont {I.}~\bibnamefont {Bazarov}},\ and\ \bibinfo {author} {\bibfnamefont {J.}~\bibnamefont {Maxson}},\ }\href@noop {} {\bibfield  {journal} {\bibinfo  {journal} {Applied Physics Letters}\ }\textbf {\bibinfo {volume} {118}} (\bibinfo {year} {2021})}\BibitemShut {NoStop}%
\bibitem [{\citenamefont {Rosenzweig}\ \emph {et~al.}(1994)\citenamefont {Rosenzweig}, \citenamefont {Barov}, \citenamefont {Hartman}, \citenamefont {Hogan}, \citenamefont {Park}, \citenamefont {Pellegrini}, \citenamefont {Travish}, \citenamefont {Zhang}, \citenamefont {Davis}, \citenamefont {Hairapetian},\ and\ \citenamefont {Joshi}}]{ROSENZWEIG1994379}%
  \BibitemOpen
  \bibfield  {author} {\bibinfo {author} {\bibfnamefont {J.}~\bibnamefont {Rosenzweig}}, \bibinfo {author} {\bibfnamefont {N.}~\bibnamefont {Barov}}, \bibinfo {author} {\bibfnamefont {S.}~\bibnamefont {Hartman}}, \bibinfo {author} {\bibfnamefont {M.}~\bibnamefont {Hogan}}, \bibinfo {author} {\bibfnamefont {S.}~\bibnamefont {Park}}, \bibinfo {author} {\bibfnamefont {C.}~\bibnamefont {Pellegrini}}, \bibinfo {author} {\bibfnamefont {G.}~\bibnamefont {Travish}}, \bibinfo {author} {\bibfnamefont {R.}~\bibnamefont {Zhang}}, \bibinfo {author} {\bibfnamefont {P.}~\bibnamefont {Davis}}, \bibinfo {author} {\bibfnamefont {G.}~\bibnamefont {Hairapetian}},\ and\ \bibinfo {author} {\bibfnamefont {C.}~\bibnamefont {Joshi}},\ }\href {https://doi.org/https://doi.org/10.1016/0168-9002(94)90387-5} {\bibfield  {journal} {\bibinfo  {journal} {Nuclear Instruments and Methods in Physics Research Section A: Accelerators, Spectrometers, Detectors and Associated Equipment}\ }\textbf {\bibinfo {volume} {341}},\ \bibinfo {pages} {379}
  (\bibinfo {year} {1994})}\BibitemShut {NoStop}%
\bibitem [{\citenamefont {Rosenzweig}\ \emph {et~al.}(2019)\citenamefont {Rosenzweig}, \citenamefont {Cahill}, \citenamefont {Dolgashev}, \citenamefont {Emma}, \citenamefont {Fukasawa}, \citenamefont {Li}, \citenamefont {Limborg}, \citenamefont {Maxson}, \citenamefont {Musumeci}, \citenamefont {Nause}, \citenamefont {Pakter}, \citenamefont {Pompili}, \citenamefont {Roussel}, \citenamefont {Spataro},\ and\ \citenamefont {Tantawi}}]{PhysRevAccelBeams.22.023403}%
  \BibitemOpen
  \bibfield  {author} {\bibinfo {author} {\bibfnamefont {J.~B.}\ \bibnamefont {Rosenzweig}}, \bibinfo {author} {\bibfnamefont {A.}~\bibnamefont {Cahill}}, \bibinfo {author} {\bibfnamefont {V.}~\bibnamefont {Dolgashev}}, \bibinfo {author} {\bibfnamefont {C.}~\bibnamefont {Emma}}, \bibinfo {author} {\bibfnamefont {A.}~\bibnamefont {Fukasawa}}, \bibinfo {author} {\bibfnamefont {R.}~\bibnamefont {Li}}, \bibinfo {author} {\bibfnamefont {C.}~\bibnamefont {Limborg}}, \bibinfo {author} {\bibfnamefont {J.}~\bibnamefont {Maxson}}, \bibinfo {author} {\bibfnamefont {P.}~\bibnamefont {Musumeci}}, \bibinfo {author} {\bibfnamefont {A.}~\bibnamefont {Nause}}, \bibinfo {author} {\bibfnamefont {R.}~\bibnamefont {Pakter}}, \bibinfo {author} {\bibfnamefont {R.}~\bibnamefont {Pompili}}, \bibinfo {author} {\bibfnamefont {R.}~\bibnamefont {Roussel}}, \bibinfo {author} {\bibfnamefont {B.}~\bibnamefont {Spataro}},\ and\ \bibinfo {author} {\bibfnamefont {S.}~\bibnamefont {Tantawi}},\ }\href
  {https://doi.org/10.1103/PhysRevAccelBeams.22.023403} {\bibfield  {journal} {\bibinfo  {journal} {Phys. Rev. Accel. Beams}\ }\textbf {\bibinfo {volume} {22}},\ \bibinfo {pages} {023403} (\bibinfo {year} {2019})}\BibitemShut {NoStop}%
\bibitem [{\citenamefont {Limborg-Deprey}\ \emph {et~al.}(2016)\citenamefont {Limborg-Deprey}, \citenamefont {Adolphsen}, \citenamefont {McCormick}, \citenamefont {Dunning}, \citenamefont {Jobe}, \citenamefont {Li}, \citenamefont {Raubenheimer}, \citenamefont {Vrielink}, \citenamefont {Vecchione}, \citenamefont {Wang},\ and\ \citenamefont {Weathersby}}]{PhysRevAccelBeams.19.053401}%
  \BibitemOpen
  \bibfield  {author} {\bibinfo {author} {\bibfnamefont {C.}~\bibnamefont {Limborg-Deprey}}, \bibinfo {author} {\bibfnamefont {C.}~\bibnamefont {Adolphsen}}, \bibinfo {author} {\bibfnamefont {D.}~\bibnamefont {McCormick}}, \bibinfo {author} {\bibfnamefont {M.}~\bibnamefont {Dunning}}, \bibinfo {author} {\bibfnamefont {K.}~\bibnamefont {Jobe}}, \bibinfo {author} {\bibfnamefont {H.}~\bibnamefont {Li}}, \bibinfo {author} {\bibfnamefont {T.}~\bibnamefont {Raubenheimer}}, \bibinfo {author} {\bibfnamefont {A.}~\bibnamefont {Vrielink}}, \bibinfo {author} {\bibfnamefont {T.}~\bibnamefont {Vecchione}}, \bibinfo {author} {\bibfnamefont {F.}~\bibnamefont {Wang}},\ and\ \bibinfo {author} {\bibfnamefont {S.}~\bibnamefont {Weathersby}},\ }\href {https://doi.org/10.1103/PhysRevAccelBeams.19.053401} {\bibfield  {journal} {\bibinfo  {journal} {Phys. Rev. Accel. Beams}\ }\textbf {\bibinfo {volume} {19}},\ \bibinfo {pages} {053401} (\bibinfo {year} {2016})}\BibitemShut {NoStop}%
\bibitem [{\citenamefont {Tan}\ \emph {et~al.}(2022)\citenamefont {Tan}, \citenamefont {Antipov}, \citenamefont {Doran}, \citenamefont {Ha}, \citenamefont {Jing}, \citenamefont {Knight}, \citenamefont {Kuzikov}, \citenamefont {Liu}, \citenamefont {Lu}, \citenamefont {Piot}, \citenamefont {Power}, \citenamefont {Shao}, \citenamefont {Whiteford},\ and\ \citenamefont {Wisniewski}}]{PhysRevAccelBeams.25.083402}%
  \BibitemOpen
  \bibfield  {author} {\bibinfo {author} {\bibfnamefont {W.~H.}\ \bibnamefont {Tan}}, \bibinfo {author} {\bibfnamefont {S.}~\bibnamefont {Antipov}}, \bibinfo {author} {\bibfnamefont {D.~S.}\ \bibnamefont {Doran}}, \bibinfo {author} {\bibfnamefont {G.}~\bibnamefont {Ha}}, \bibinfo {author} {\bibfnamefont {C.}~\bibnamefont {Jing}}, \bibinfo {author} {\bibfnamefont {E.}~\bibnamefont {Knight}}, \bibinfo {author} {\bibfnamefont {S.}~\bibnamefont {Kuzikov}}, \bibinfo {author} {\bibfnamefont {W.}~\bibnamefont {Liu}}, \bibinfo {author} {\bibfnamefont {X.}~\bibnamefont {Lu}}, \bibinfo {author} {\bibfnamefont {P.}~\bibnamefont {Piot}}, \bibinfo {author} {\bibfnamefont {J.~G.}\ \bibnamefont {Power}}, \bibinfo {author} {\bibfnamefont {J.}~\bibnamefont {Shao}}, \bibinfo {author} {\bibfnamefont {C.}~\bibnamefont {Whiteford}},\ and\ \bibinfo {author} {\bibfnamefont {E.~E.}\ \bibnamefont {Wisniewski}},\ }\href {https://doi.org/10.1103/PhysRevAccelBeams.25.083402} {\bibfield  {journal} {\bibinfo  {journal} {Phys. Rev. Accel.
  Beams}\ }\textbf {\bibinfo {volume} {25}},\ \bibinfo {pages} {083402} (\bibinfo {year} {2022})}\BibitemShut {NoStop}%
\end{thebibliography}%

\end{document}